\documentclass{article}
\usepackage{amsmath}
\usepackage{amsfonts}
\usepackage{mathrsfs}  
\usepackage{amssymb}
\usepackage{graphicx}
\usepackage{cancel}
\usepackage{soul}
\usepackage{ulem}

\usepackage{titlesec}

\titleformat{\paragraph}
{\normalfont\normalsize\bfseries}{\theparagraph}{1em}{}
\titlespacing*{\paragraph}
{0pt}{3.25ex plus 1ex minus .2ex}{1.5ex plus .2ex}

\usepackage{xcolor}
\usepackage{tensor}

\definecolor{blue}{rgb}{0,0,1}
\definecolor{green}{rgb}{0,0.8,0.3}
\definecolor{turkos}{rgb}{0,0.4,0.4}
\definecolor{red}{rgb}{1,0,0}
\definecolor{darkblue}{rgb}{0,0,0.5}
\definecolor{darkgreen}{cmyk}{1,0,1,0}
\definecolor{green0}{rgb}{0,1,0}
\newcommand{\pdv}[2]{\frac{\partial{#1}}{\partial{#2}}}
\newcommand{\dd}[1]{\hbox{d}{#1}}
\newcommand{\vb}[1]{\hbox{\bf{#1}}}

\begin{document}

\title{The near equilibrium Einstein-Boltzmann system with a simplified collision term}
\author{P. Semr\'en\footnote{Department of Physics, University of Ume{\aa}, Sweden, \\Email: philip.semren@umu.se, michael.bradley@umu.se} ,  M. Bradley$^{*}$, J.M.S. Oliveira\footnote{Centre of Mathematics, University of Minho, Portugal,\\ Email: jmiguel.oliveira17@gmail.com, mpr@math.uminho.pt} , M.P. Machado Ramos$^{\dagger}$}
\maketitle

\begin{abstract}
A simplified relativistic kinetic theory for gases with internal degrees of  freedom, based on a BGK-type collision term, is considered. First the Boltzmann equation is rewritten in tetrad form and then 
thermal coefficients are determined to first order in the Chapman-Enskog expansion 
for general spacetimes.
The results are used to construct a self-consistent system of first order differential equations, equivalent to the Einstein-Boltzmann system, 
for some spatially homogeneous models 
with viscosity and heat flow. 
\end{abstract}

\section{Introduction}\label{introduction}

Due to their complexity, few self-consistent solutions to the combined system of  Einstein's equations and Boltzmann's equations are known or likely to be found in terms of elementary functions or quadratures. Hence it is of interest to reduce this system to a manifest integrable one, which could be further investigated numerically. As a first step, we will in this paper examplify such a procedure on a spatially homogeneous system.

The development of a relativistic kinetic theory started with J\"{u}ttner's  generalization of the Maxwell distribution function for a relativistic gas \cite{Juttner}, followed by the generalization of the Boltzmann equation to a covariant formulation by Lichnerowicz and Marrot \cite{Marrot}.
Other important contributions were the works by Israel \cite{Israel} and Kelly \cite{Kelly}, where relativistic versions of the Navier-Stokes and Fourier laws, together with the transport coefficients, were obtained from the relativistic Boltzmann equation for a simple gas using the Chapman-Enskog method \cite{Cerci,ChapmanEnskog}. Furthermore, it was recognized in \cite{Israel} that a simple gas has a bulk viscosity, a purely relativistic effect. In {\cite{Israel} also conservation laws and the relativistic version of the H-theorem are presented and the notion of the state of thermodynamical equilibrium in a gravitational field is clarified. 

Historically, by considering first-order deviations from thermodynamic equilibrium --- either through phenomenological or kinetic arguments --- the resulting transport equations have generally been found to be acausal and plagued by instabilities. This is for instance true for the standard Eckart theory {\cite{Eckart}. However, with the rise of the Bemfica-Disconzi-Noronha-Kovtun (BDNK) formalism \cite{BDNK1,BDNK2}, providing a phenomenological approach to causal first-order relativistic thermodynamics, there have recently been renewed interest in microscopic descriptions of first-order thermodynamics through kinetic theory \cite{Rocha, Garcia-Perciante1,Garcia-Perciante2}.

Since the usual binary collision term in the Boltzmann equation is given by a complicated integral over momenta, different simplified models have been used in the literature. One of these, which is based on a single constant relaxation time, is the so called BGK model, suggested by Bhatnagar, Gross and Krook \cite{BGK}. Relativistic generalisations of the BGK-type collision terms are given by, e.g., Marle \cite{Marle1,Marle2}, Anderson and Witting \cite{AndersonWitting} and Pennisi and Ruggeri \cite{PennisiRuggeri2}.

In several works by Kremer  the relativistic Bolztmann equation is applied to the study of relativistic monoatomic gases in homogeneous and isotropic universes and in Schwarzschild spacetime, \cite{Kremer1, Kremer2, Kremer3}.  A relativistic extended thermodynamics theory of rarefied polyatomic gases was established by Pennisi and Ruggeri in \cite{PennisiRuggeri1}, and in \cite{PennisiRuggeri2} a relativistic BGK-type model was proposed. In a work by Oliveira, Ramos and Soares \cite{ORS} this theory, together with the Chapman-Enskog method, was used to derive the constitutive equation for the non-equilibrium dynamical pressure in the case of a rariefied relativistic polyatomic gas in the Robertson-Walker spacetimes.

One of the aims of the present work is to construct the energy-momentum tensor in tetrad form for a general spacetime in first order deviations from equilibrium. This is in constrast to previous works, where specific spacetimes have been considered.
To this end we use the simplified relativistic BGK-type kinetic theory model presented in \cite{PennisiRuggeri2} above, which generalizes the collision term to a form suitable for using
an Eckart frame, together with its extention to polyatomic gases \cite{PennisiRuggeri1}. 
Another goal is to illustrate  how one may reduce a combined Einstein-Boltzmann system to a self-consistent set of first order differential equations. To obtain a relatively simple system we will consider some cosmological models where all types of terms in the energy-momentum for an imperfect fluid appear, but the Einstein-Boltzmann systems are given by a set of ordinary differential equations in time.

First we rewrite the Boltzmann equation in tetrad form, using a frame comoving with the fluid. In this way the integrations with respect to the momenta
 become the same as in special relativity, and hence are independent of the metric.

Then, on using the Chapman-Enskog expansion, cf., e.g., \cite{ChapmanEnskog}, to first order, a near equilibrium configuration, consistent with the Eckart theory, \cite{Eckart}, is constructed. This is used to calculate the energy-momentum tensor, from which the bulk and shear viscosity coefficients are read off, as well as to determine the particle current density, from which the coefficient of heat conductivity is found. 

The so constructed energy-momentum tensor is then used to find self-consistent solutions to the Einstein-Boltzmann system with viscosity and heat flow. For this we consider a class of homogeneous and tilted LRS (Locally Rotationally Symmetric) models of Bianchi type VIII, and hence extend the earlier work, \cite{ORS}, where Robertson-Walker cosmologies were studied. The equations are rewritten as an integrable system of first order ordinary differential equations, suitable for numerical integration. The time evolution of some of the models are then studied.

The paper is organized in the following way: in Section \ref{preliminaries} some preliminaries of the kinetic theory are given, and  
in Section \ref{sectetrad} the description in a tetrad comoving with the fluid is treated.   
In Section \ref{secChapmanEnskog} the Chapman-Enskog method is used to find the distribution function to first order in deviations from thermodynamical
equilibrium. Then, in Section \ref{secThermocoeff}, the first and second moments of the distribution funcction are determined together with the thermodynamical coefficients.
In Section \ref{secLRS} the cosmological model is presented, and the corresponding
Einstein-Boltzmann system is constructed. Then, in section \ref{secNumerical} this system is solved numerically, and finally the results are discussed in section \ref{discussion}.

\section{Preliminaries}\label{preliminaries}

We consider a kinetic model for a relativistic and classical polyatomic gas, based on a BGK-type \cite{PennisiRuggeri1,BGK} collision term,  in a space-time $(M, g_{\mu\nu})$
characterized by a four-dimensional differentiable manifold $M$ endowed with a Lorentzian metric $g_{\mu\nu}$ of signature -2.
The following index conventions will be used:
for spacetime coordinates $x^\mu$ (where $x^0\equiv ct$) $\mu,\nu,...=0,1,2,3$,
for spatial coordinates $x^\alpha$, $\alpha,\beta...=1,2,3$, for tetrad indices $i,j,...=0,1,2,3$
and for spatial tetrad indices $a,b,...=1,2,3$.

The gas is supposed to be self-gravitating, and hence
these solutions satisfy the Einstein-Boltzmann system, consisting of  Einstein's equations 
\begin{equation}
G_{\mu\nu}=\frac{8\pi G}{c^4}T_{\mu\nu}+\Lambda g_{\mu\nu},
\end{equation}
where $G_{\mu\nu}$ is the Einstein tensor, $G$ is the gravitational constant, $\Lambda$ is the cosmological constant, $c$ is the speed of light and $T_{\mu\nu}$ is the energy momentum tensor, and of the relativistic Boltzmann equation
\begin{equation}\label{Boltzmann0}
p^\mu\left(\frac{\partial f}{\partial x^\mu}-\Gamma^\alpha_{\; \mu\nu}p^\nu\frac{\partial f}{\partial p^\alpha}\right)=Q(f,f) \, ,
\end{equation}
where $p^\mu$ is the momentum four-vector, $\Gamma^\alpha_{\; \mu\nu}$ are the Christoffel symbols and $Q(f,f)$ is the collision term.
Here $f= f(t, {x^\alpha}, {p^\alpha}, \cal{I})$ is the extended distribution function, where the additional variable $\cal I$ represents the continuous internal energy of the molecules due to the internal degrees of freedom.

 The particle current density, $V^{\mu}$, the energy-momentum tensor, and the entropy 4-vector are obtained from the distribution function $f$ through 
\begin{equation}\label{Vmu}
V^{\mu}[f]= mc\int^{\infty}_0\int_{R^3}fp^\mu \Phi({\cal{I}})d{\bf{P}}d{\cal{I}}\, .
\end{equation}
\begin{equation}\label{Tmunu}
T^{\mu\nu}[f]=\frac{1}{mc}\int^{\infty}_0\int_{R^3}\left(mc^2+{\cal{I}}\right)fp^\mu p^\nu \Phi({\cal{I}})d{\bf{P}}d{\cal{I}}\, ,
\end{equation}
and
\begin{equation}\label{hmu}
h^\mu[f] = -kc\int^{+\infty}_0\int_{R^3}f\ln(f)p^\mu \Phi({\mathcal{I}})\dd{\vb{P}}d{\cal{I}} \, ,
\end{equation}
respectively. Here $m$ is the particle rest mass, $\Phi({\cal{I}})$ is the density of state of the internal degrees of freedom, $d{\bf{P}}$ is the invariant momentum measure, which is simplest expressed in tetrad components, where it takes its special relativistic form $d^3{\bf{p}}/p_0$
with  $p_0=\sqrt{m^2c^2-p_a p^a}$. The moments $V^\mu$ and $T^{\mu\nu}$ are required to satisfy the conservation laws $V^\mu_{\; ;\mu}=0$ and $T^{\mu\nu}_{\;\;\; ;\nu}=0$, respectively, if $f$
is a solution of the relativistic Boltzmann equation (\ref{Boltzmann0}). Additionally, the entropy should be increasing,  $h^\mu_{\; ;\mu}\geq 0$ in accordance with the H-theorem. Using the Boltzmann equation, this holds provided that 
\begin{equation}\label{Qrelation1}
\int^{\infty}_0\int_{R^3} Q(f,f) \Phi({\cal{I}})d{\bf{P}}d{\cal{I}} = 0\,,  
\end{equation}
\begin{equation}\label{Qrelation2}
 \int^{\infty}_0\int_{R^3}\left(mc^2+{\cal{I}}\right)p^\mu Q(f,f) \Phi({\cal{I}})d{\bf{P}}d{\cal{I}}= 0\, , 
\end{equation}
and that 
\begin{equation}\label{Qrelation3}
 -kc\int^{+\infty}_0\int_{R^3} Q(f,f)\ln(f)\Phi({\mathcal{I}})\dd{\vb{P}} d{\cal{I}}  \geq 0 \, .
\end{equation}
When considering simplified collision terms, these relations should still be satisfied and hence restrict the space of physically acceptable models.

The Boltzmann equation (\ref{Boltzmann0}) can be simplified using a BGK model \cite{BGK}, in which the collision term $Q$ is
given by an approximation, based on a single constant relaxation time. In this paper 
we consider the BGK type model for a relativistic polyatomic gas developed in \cite{PennisiRuggeri1,PennisiRuggeri2} with collision term  
\begin{equation}\label{Qf}
Q(f)=\frac{u_\mu p^\mu}{c^2\tau}\left(f_{EP}-f-\frac{\gamma^\star p^\nu q_\nu}{pmc^2}\frac{A(\gamma)}{B(\gamma)}f_{EP}\right)\, ,
\end{equation}
where the equilibrium distribution function  $f_{EP}$ is the natural generalization to polyatomic gases of the J\"uttner equilibrium distribution function for monatomic gases, 
$f_{EM}(t,x^\alpha,p^\alpha)$, see \cite{Juttner}, and is given by
\begin{equation}\label{fep}
f_{EP}(t,x^\alpha,p^ \alpha,{\cal{I}})=\frac{n}{4\pi m^2 c k T K_2(\gamma)A(\gamma)}\exp\left(-\frac{\gamma^\star}{\gamma}\frac{u_\mu p^\mu}{kT}\right) .
\end{equation}
This specific collision term is conditional on using the Eckart thermodynamic frame, given by the relations
\begin{equation}
V^{\mu}[f] = V^{\mu}[f_{EP}]\,, \quad T^{\mu\nu}[f]u_\mu u_\nu =  T^{\mu\nu}[f_{EP}]u_\mu u_\nu \, ,
\end{equation} 
since it is only then that equations (\ref{Qrelation1})--(\ref{Qrelation3}) are satisfied. This follows directly from the special-relativistic result in \cite{PennisiRuggeri2} on using a freely-falling coordinate system or a tetrad description in equations (\ref{Qrelation1})--(\ref{Qrelation3}).
Note that $u^\mu$ is thus the Eckart four-velocity, i.e. the average 4-velocity of the gas. As for the remaining quantities appearing in the collision term and $f_{EP}$, $\tau$ is the relaxation time in the Eckart frame, 
$n$ is the number density of particles,
 $p$ is the equilibrium pressure, $q_\mu$ is the heat flow, $k$ is the Boltzmann constant, $T$ is the temperature of the gas and
 \begin{equation}\label{gammastar}
\gamma^\star\equiv\gamma\left(1+\frac{{\cal{I}}}{mc^2}\right) \quad \hbox{with} \quad  \gamma\equiv\frac{mc^2}{kT}\, .
 \end{equation}
 $K_2(\gamma)$ represents the modified Bessel function of second kind, 
defined by
\begin{equation}
K_n(\gamma) \equiv \left(\frac{\gamma}{2}\right)^n \frac{\Gamma(1/2)}{\Gamma(n+1/2)}
\int_{1}^{\infty} e^{-\gamma y}(y^2-1)^{n-1/2} \, dy ,
\quad n=0,1,2,\ldots
\label{BF}
\end{equation} 
and the integrals $A$ and $B$ are
\begin{equation}\label{intA}
A\equiv A(\gamma) \equiv \frac{\gamma}{K_2(\gamma)} \int_0^{\infty}\frac{K_2(\gamma^\star)}{\gamma^\star} \,
\Phi({\cal I}) d{\cal I},   \qquad \hbox{and}
\end{equation}
\begin{equation}\label{intB}
B\equiv B (\gamma)\equiv \frac{\gamma}{K_2(\gamma)}\int_0^{\infty}{K_3(\gamma^\star)}\Phi({\cal I}) d{\cal I} \, ,
\end{equation}
respectively.

On substituting the equilibrium distribution (\ref{fep}) into equation (\ref{Tmunu}), one obtains the following equilibrium energy-momentum tensor
\begin{equation}\label{Tmunueq}
T\indices{^\mu^\nu}[f_{EP}] = \frac{\mu+p}{c^2}u^\mu u^\nu -p g\indices{^\mu^\nu}\, ,
\end{equation}
where the equilibrium pressure and energy density are given by
\begin{equation}\label{mupeq}
p=nkT\, , \quad \hbox{and} \quad \mu=p\left[\frac{B}{A}-1\right]\, ,
\end{equation}
respectively. Inserting the full distribution in the Eckart frame, we get
\begin{equation}\label{energymomentum}
T\indices{^\mu^\nu}[f] = \frac{\mu+p + \Pi}{c^2}u^\mu u^\nu-\left(p+\Pi\right) g\indices{^\mu^\nu} + \frac{2}{c^2}q^{(\mu}u^{\nu)} + \pi\indices{^\mu^\nu}\, ,
\end{equation}
where $p$ and $\mu$ are given by the equilibrium expressions, $\Pi$ is the bulk viscous pressure, and $\pi\indices{^\mu^\nu}$ is the anisotropic pressure. Finally, the number current density is 
\begin{equation}
V^{\mu}[f] = V^{\mu}[f_{EP}] =m nu^\mu\, .
\end{equation}
The aim of this paper is now to explicitly calculate the dissipative contributions $q^\mu$, $\Pi$, and $\pi\indices{^\mu^\nu}$ using the kinetic Chapman-Enskog method, considering small deviations away from the equilibrium contribution $f_{EP}$. 

For the models considered in section \ref{secLRS}, we will then assume that the density of the internal energy states may be approximated with the polytropic form $\Phi({\cal{I}})\propto{\cal{I}}^\alpha$ \cite{Pavic}.
This form gives a description of the internal energies for, e.g., rarified polyatomic gases}. The exponent $\alpha$ then is related to the number of degrees of freedom, $D$, through
$\alpha=(D-5)/2$.
For example, $\alpha=0$ could correspond to a diatomic gas in an intermediate temperature interval where the two rotational degrees of freedom are exited,
whereas a monoatomic gas is obtained in the limit $\alpha \rightarrow -1$. Note however, that this description is classical,  in particular particle creation cannot occur. Hence it cannot be used beyond certain temperatures, where dissociations, ionizations and particle and photon creation would take place.

In summary, we adopt the simplified theory above with collision term $Q$, given by equation (\ref{Qf}), and internal energy density $\Phi({\cal{I}})\propto{\cal{I}}^\alpha$, as being a model which is consistent with
general covariance, gives a locally conserved energy-momentum, obeys the second law of thermodynamics and gives a fair description for a single species of polyatomic molecules 
in certain temperature intervals in the classical regime.

\section{Converting to tetrad form}\label{sectetrad} 

The first and second moments of the distribution functions, i.e. the particle current and energy-momentum tensor respectively, are obtained from integration over
the linear momenta at fixed spacetime points. Hence these parts of the calculations are unaffected by the curvature of spacetime, and can be done as in special relativity,
independently of the metric. 
Because of this we will perform our calculations using a Lorentz tetrad, and consequently
 first rewrite the Boltzmann equation, which in a coordinate basis is given by \eqref{Boltzmann0},  in tetrad form.

First we will use the freedom in coordinates to choose the 4-velocity $u^\mu$ of the distribution function $f$ to be proportional to the $t$-lines, i.e. $u^\mu\propto \frac{\partial}{\partial t}$. Then we adjust the tetrad so that one of the basis vectors is along $u^\mu$, i.e. so that $u^i=c\delta_0^i$.

On splitting the metric into time and spatial components we have
\begin{equation}
ds^2=g_{00}c^2dt^2+2g_{0\alpha}cdtdx^\alpha+g_{\alpha\beta}dx^\alpha dx^\beta  \, ,
\end{equation}
where $\alpha,\beta=1,2,3$ numbers the spatial coordinates.
We will use a Lorentz tetrad ${\omega^i}$, with tetrad metric $\eta_{ij}=\hbox{diag}(1,-1,-1,-1)$, i.e.
\begin{equation}
ds^2=\eta_{ij}\omega^i\omega^j\, ,
\end{equation}
where the basis forms and vectors are given by
\begin{equation}
\omega^i=\omega^i_{\; \mu}dx^\mu \,   \quad \hbox{and} \quad X_i=X_i^{\; \mu}\frac{\partial}{\partial x^\mu}\, ,
\end{equation}
respectively. Here the coefficients $\omega^i_{\; \mu}=\omega^i_{\; \mu}(x^\nu)$ and $ X_i^{\; \mu}=X_i^{\; \mu}(x^\nu)$ satisfy
\begin{equation}
\omega^i_{\; \mu}X_j^{\; \mu}=\delta_j^i \quad \hbox{and}\quad \omega^i_{\; \mu}X_i^{\; \nu}=\delta_\mu^\nu \, .
\end{equation}
We choose the tetrad so that
\begin{equation}
\omega^0=\sqrt{g_{00}}\left(cdt+\frac{g_{0\alpha}}{g_{00}}dx^\alpha\right)\, ,
\end{equation}
which implies that the timelike basis vector is tangent to the $t$-lines
\begin{equation}
X_0=\frac{1}{\sqrt{g_{00}}}\frac{\partial}{c\partial t}\, .
\end{equation}
In this frame $\omega^a_{\; 0}=0$ for the spatial forms $\omega^a$, $a=1,2,3$, i.e.
\begin{equation}
\omega^a\equiv\omega^a_{\; \mu}dx^\mu=\omega^a_{\; \alpha}dx^\alpha \; ,
\end{equation}
and hence it will also follow that the 3 by 3 matrices $\omega^a_{\; \alpha}$ and $X_a^{\; \alpha}$ satisfy
\begin{equation}
\omega^a_{\; \alpha}X_b^{\; \alpha}=\delta_a^b \quad \hbox{and}\quad \omega^a_{\; \alpha}X_a^{\; \beta}=\delta_\alpha^\beta \, .
\end{equation}

\subsection{The Boltzmann equation}

We now proceed to rewrite the Boltzmann equation (\ref{Boltzmann0})
in terms of the components in the tetrad introduced above. First we note that the transformation to tetrad components of the momenta, $p^a=\omega^a_{\; \alpha}(x^\mu)p^\alpha$,
changes the dependence on $x^\mu$.
On transforming the derivatives  in (\ref{Boltzmann0}) accordingly, and 
using 
\begin{equation}
\Gamma^\mu_{\; \nu\sigma}=X_i^{\; \mu}\omega^j_{\; \nu}\omega^k_{\; \sigma}\gamma^i_{\; j k}+X_i^{\; \mu}\omega^i_{\; \nu , \sigma}
\end{equation}
for the relation between the Christoffel symbols and the Ricci rotation coefficients, $\gamma^i_{\; j k}=-X_j^{\; \mu}X_k^{\; \nu}\omega^i_{\; \mu ; \nu}$ , 
straightforward calculations give
\begin{equation}\label{Boltzmanntetrad2}
p^i X_i(f)
-\gamma^a_{\; j k}p^j p^k  \frac{\partial f}{\partial p^a}=Q(f,f)\, ,
\end{equation}
where now $f=f(x^\mu,p^a)$,
for the Boltzmann equation in the tetrad basis. For an alternative way of deriving this equation, see \cite{ACGS}\footnote{The left hand side of (\ref{Boltzmanntetrad2}) agrees with the Liouville operator in equation (123) in \cite{ACGS} with the identification 
$e^\mu_{i}\hat\Gamma^{k}_{\; \mu j}=\gamma^k_{\; j i}$.}.

As said above, we choose the 4-velocity $u$ along $X_0$, so that $u^i=c\delta_0^i$ with norm $u^i u_i=c^2$.
The covariant derivative of $u$ can be written in terms of the kinematic quantities of $u^i$ as
\begin{equation}\label{kinematic}
u^i_{\; ;j}=\frac{1}{c^2}a^i u_j-\frac{1}{3}h^i_{\; j}\Theta+\sigma^i_{\; j}+\omega^i_{\; j}
\end{equation}
where
\begin{equation}
\Theta\equiv u^i_{\; ;  i}\, ,\;\; a_i\equiv u^j u_{i;j}\, ,\;\; \omega_{ij}\equiv h_i^{\; k}h_j^{\; l}u_{[k;l]}\, ,\;\; \sigma_{ij}\equiv h_i^{\; k}h_j^{\; l}\left(u_{(k;l)}+\frac{1}{3}h_{kl}\Theta\right)
\end{equation}
are the expansion, acceleration, vorticity and shear respectively, and the projection operator $h_{ij}\equiv \frac{1}{c^2}u_i u_ j - \eta_{ij}$\; .
The covariant derivative of $u^i$ can also be written as
\begin{equation}
u^i_{\; ;j}=X_j(c\delta^i_0)+\gamma^i_{\; kj}c\delta^k_0=c\gamma^i_{\; 0j}\, ,
\end{equation}
so that
\begin{equation}\label{kingam}
\gamma^a_{\; 00}= \frac{1}{c^2}a^a\; ,\quad \gamma^a_{\; 0b}=\frac{1}{3c}\Theta\delta^a_{\; b}+\frac{1}{c}\left(\sigma^a_{\; b}+\omega^a_{\; b}\right)\, .
\end{equation}

The 2:nd term on the left hand side of the Boltzmann equation (\ref{Boltzmanntetrad2}) may then be expanded as
\begin{eqnarray}\label{BLHS2}\nonumber
-\gamma^a_{\; j k}p^j p^k  \frac{\partial f}{\partial p^a}
&=&-\frac{1}{c^2}a^a(p^0)^2\frac{\partial f}{\partial p^a}-\frac{1}{3c}\Theta p^0 p^a \frac{\partial f}{\partial p^a}-\frac{1}{c}p^0 p^b \sigma^a_{\; b} \frac{\partial f}{\partial p^a}
\\ 
&&-\frac{1}{c}p^0 p^b \omega^a_{\; b} \frac{\partial f}{\partial p^a}-\gamma^a_{\; b 0}p^b p^0  \frac{\partial f}{\partial p^a}-\gamma^a_{\; b c}p^b p^c  \frac{\partial f}{\partial p^a}\, ,
\end{eqnarray}
where the zeroth component, $p^0$, of the 4-momentum is given by
\begin{equation}\label{eqp0}
p^0=p_0=\sqrt{m^2c^2-p_b p^b}
\end{equation}
in terms of the independent variables $p^a$, since $p_i p^i=m^2 c^2$.

\subsection{Conservation laws}\label{secCons}

In general, the energy-momentum tensor can be written in tetrad form as
\begin{equation}\label{energymomentum}
T\indices{^i^j} = \frac{\mu}{c^2}u^iu^j +\left(p+\Pi\right)h\indices{^i^j} + \frac{2}{c^2}q^{(i}u^{j)} + \pi\indices{^i^j}
\end{equation} Then, from the twice contracted Bianchi identies and Einstein's equations, we also have
\begin{equation}\label{Bianchi1}
0=u_i\nabla_j T^{ij}=cX_0(\mu)+\left(\mu+p+\Pi\right)\theta-\frac{1}{c^2}q^ia_i+q^i_{\; ;i}-\pi^{ij}\sigma_{ij}
\end{equation}
and
\begin{eqnarray}\nonumber
0=h^{k}_{\; i}\nabla_j T^{ij}&=&h^k_{\; j}\left(\frac{1}{c}X_0(q^j)+\pi^{jl}_{\;\; ;l}\right)+h^{kj}X_j\left(p+\Pi\right)-\frac{4}{3c^2}\theta q^k\\\label{Bianchi2}
&&-\frac{1}{c^2}\left(\sigma^k_{\; j}+\omega^k_{\; j}\right)q^j-\frac{1}{c^2}\left(\mu+p+\Pi\right)a^k\, ,
\end{eqnarray}
as integrability conditions. The distribution function $f$ is given in terms of the two thermodynamic variables particle density, $n$, and temperature, $T$. So to complete the system we add the particle conservation equation
\begin{equation}\label{ndens}
cX_0(n)+n\theta=0 \; ,
\end{equation}
which follows from $V^i_{\, ;i}=0$ and the use of a particle frame.
One may also obtain an evolution equation for the temperature.
Differentiation of the energy density $\mu$, Eq. (\ref{mupeq}),  gives
\begin{equation}\label{mudot}
cX_0(\mu)=-\theta\mu+pc\frac{X_0(T)}{T}\frac{c_v}{k}\, ,
\end{equation}
where the specific heat per particle, $c_v$, is given by 
\begin{equation}\label{cv}
c_v=\frac{\partial}{\partial T}\left(\frac{\mu}{n}\right)=k\left[-1+5\frac{B}{A}-\frac{B^2}{A^2}+\gamma\frac{C}{A}\right] \, ,
\end{equation}
with
\begin{equation}\label{eqC}
C\equiv C(\gamma) \equiv \frac{1}{K_2(\gamma)}\int_0^\infty \gamma^\star K_2(\gamma^\star)\Phi(\mathcal{I})d\mathcal{I}\, .
\end{equation}
Substitution of (\ref{mudot}) into (\ref{Bianchi1}) then gives
\begin{eqnarray}\nonumber
c\frac{X_0(T)}{T}&=&-\theta\frac{k}{c_v}-\frac{1}{T(\partial\mu/\partial T)}_n\left[\theta\Pi+q^i_{\;\; ;i}-\frac{1}{c^2}a_i q^i - \sigma_{ij}\pi^{ij}\right]\\\label{Ttimeder}
&=&-\frac{k}{pc_v}\left[\theta(p+\Pi)+q^i_{\;\; ;i}-\frac{1}{c^2}a_i q^i - \sigma_{ij}\pi^{ij}\right]\, ,
\end{eqnarray}
which agrees with Eq. (2.36) in \cite{Marteens} (note that their signature convention is $+2$).

\section{Chapman-Enskog expansion}\label{secChapmanEnskog}

We now apply the Chapman-Enskog method, \cite{ChapmanEnskog}, to obtain 
the non-equilibrium distribution function as an approximation to the solution of the Boltzmann equation \eqref{Boltzmann0}, considering a polyatomic gas with equilibrium distribution given by (\ref{fep}), in a general spacetime.
This procedure consists first, in assuming that the distribution function $f$ is expanded as a perturbation of the equilibrium distribution function $f_{_{EP}}$, of the form
\begin{equation}\label{fansatz2}
f=f_{EP}(1+\epsilon \phi_P)
\end{equation}
where $\phi_{_P}$ is a small quantity for processes close to equilibrium and $\epsilon$ has been introduced to indicate terms of order one in the relaxation time $\tau$. Then, we introduce the perturbed distribution function into the BGK model with collision term given by equation \eqref{Qf}, which becomes 
\begin{equation}\label{collision2}
Q=\frac{p_0}{c\tau\epsilon}\left(f_{EP}-f-\epsilon\frac{\gamma^\star p^a q_a}{pmc^2}\frac{A(\gamma)}{B(\gamma)}f_{EP}\right)\, .
\end{equation}
Substitution of (\ref{fansatz2}) and (\ref{collision2}) into (\ref{Boltzmanntetrad2}) then gives
\begin{equation}\label{phiP0}
\phi_P=-\frac{c\tau}{p_0}\left(p^i X_i(f_{EP})-\gamma^a_{\; j k}p^j p^k  \frac{\partial f_{EP}}{\partial p^a}\right)
-\frac{\gamma^\star p^a q_a}{pmc^2}\frac{A(\gamma)}{B(\gamma)} 
\end{equation}
for $\phi_P$ to lowest order in $\epsilon$. Note that no derivatives of $\phi_P$ appear to this order. However, the expression does involve an integral over $\phi_P$ through the heat flow
\begin{equation}
q^a = cT\indices{^0^a}[f_{EP}\phi_P] =  \frac{1}{m}\int^{\infty}_0\int_{R^3}\left(mc^2+{\cal{I}}\right)f_{EP}\phi_Pp^0 p^a \Phi({\cal{I}})d{\bf{P}}d{\cal{I}}\, ,
\end{equation}
and is hence only an implicit solution for $\phi_P$. 
Substitution of equation (\ref{BLHS2}) and the expression (\ref{fep}) for $f_{EP}$ 
gives
\begin{eqnarray}\label{phiP3}
\phi_P=-\frac{c\tau}{p_0}\left(p^i\frac{X_i(n)}{n}+p^i\frac{X_i(T)}{T}\left(1-\frac{B}{A}+\frac{p_0\gamma^\star}{mc}\right)\right)\\\nonumber
+\frac{c\tau}{p_0}\frac{\gamma^\star }{mc^2}\left(
\frac{1}{3}\Theta p_a  p^a+\frac{1}{c}p_a a^a p_0 + \sigma_{a b} p^a p^b \right)-p^a q_a\frac{\gamma^\star}{pmc^2}\frac{A}{B} \, ,
\end{eqnarray}
where the vorticity term disappears since $\omega_{ab}=-\omega_{ba}$, the acceleration term vanishes since $a^a p_a=0$  and $p^a p^b \gamma_{abc}=0$ since $\gamma_{abc}=-\gamma_{bac}$. The identity
\begin{equation}
(K_2 A)_{,\gamma}\gamma=(2A-B)K_2
\end{equation} was also used.

On using the particle conservation equation (\ref{ndens})
and equation (\ref{mupeq}) for the equilbrium pressure,
together with the twice contracted Bianchi identity (\ref{Bianchi2})
\begin{equation}
X_a(p)=\frac{1}{c^2}(\mu+p)a_a\, ,
\end{equation}
taken at equilbrium since (\ref{phiP3}) is already of first order,
and the evolution equation for the temperature, (\ref{Ttimeder})
\begin{equation}
cX_0(T)+\frac{p}{nc_v}\theta=0 \, ,
\end{equation}
also at equilbrium,
we are able to rewrite \eqref{phiP3} as
\begin{eqnarray}\nonumber
\phi_P&=&\left[\frac{c\tau m\gamma^\star}{3p_0}\left(1+\frac{3 p_0}{\gamma^\star mc}-\frac{p_0^2}{m^2 c^2}\right)+\frac{\tau k}{c_v}\left(\frac{p_0\gamma^\star}{mc}+1-\frac{B}{A}\right)\right]\theta\\\nonumber
&&+\frac{c\tau\gamma^\star}{p_0 mc^2}p^a p^b \sigma_{ab}-\frac{c\tau}{p_0 T}\left(\frac{p_0\gamma^\star}{mc}-\frac{B}{A}\right)p^a\left(X_a(T)-\frac{1}{c^2}Ta_a\right)\\
&&-p^a q_a\frac{\gamma^\star}{pmc^2}\frac{A}{B}\, .
\end{eqnarray}

\section{Thermodynamic coefficients}\label{secThermocoeff}

In this section we present the first and second moments of the distribution function, which are given by equations (\ref{Vmu}) and (\ref{Tmunu}), and from them extract
the thermodynamic coefficients. They are the heat conductivity $\kappa$, bulk viscosity $\zeta$ and shear viscosity $\eta$.

\subsection{Particle current density}

It turns out that, due to the implicit nature of the solution for $\phi_P$, the relation for the heat flow $q_a$ is obtained from the particle current density rather than the energy-momentum tensor. Calculating the corresponding component of the energy-momentum tensor only gives $q^a=q^a$. However, by using the Eckart condition
\begin{equation}
V^i[f] - V^i[f_{EP}] = 0\,,  
\end{equation}
we get
\begin{equation}
q_a = \kappa\left( X_a(T) - \frac{1}{c^2}Ta_a \right)\,, \label{eq:qeq}
\end{equation}
where 
\begin{align}\label{kappa}
\kappa &= \frac{\tau kp}{3m}\gamma^3\frac{B}{A}\left(\frac{\Gamma B}{\gamma A^2} - \frac{3}{\gamma^2}\right)\, ,\quad\hbox{with} \\\label{Gamma}
\Gamma & \equiv \frac{1}{K_2(\gamma)}\int K_2(\gamma^*)\left(\frac{1}{\gamma^*}-\frac{K_1(\gamma^*)}{K_2(\gamma^*)} + \frac{Ki_{1}(\gamma^*) }{K_2(\gamma^*)}\right)\Phi(\mathcal{I}) \dd{\mathcal{I}}\, .
\end{align}
Here the Bickley-Naylor functions, or modified Bessel integrals, are defined through \cite{GargantiniPomentale}
\begin{equation}\label{BickleyNaylor}
Ki_n(\gamma)\equiv \int_1^\infty\frac{e^{-\gamma t}dt}{t^n\sqrt{t^2-1}} \, .
\end{equation}

\subsection{Energy momentum tensor}

On calculating the components of the energy momentum tensor, using the expression for $\phi_P$, and comparing with the general form (\ref{energymomentum}) of the energy-momentum tensor, we get the following dissipative additions. 

The anisotropic pressure $\pi\indices{_i_j}$ 
is given by
\begin{align}
\pi\indices{_a_b} &= 2\eta \sigma\indices{_a_b} \, ,\quad\hbox{where} \label{eq:pieq}\\
\eta &= \frac{\tau p \gamma}{15}\left(\frac{3B}{\gamma A} - \frac{C-D+E}{A}\right)\, , \label{eta}
\end{align}
while the bulk viscous pressure $\Pi$ 
is given by
\begin{align}
\Pi &= - \zeta \Theta \, ,\quad\hbox{where}  \label{eq:Pieq}\\
\zeta & =- p\tau\left(\frac{5BA-B^2+\gamma C A}{5BA-B^2+\gamma CA - A^2}\right)+\frac{5}{3}\eta=-p\tau\frac{c_v+k}{c_v}+\frac{5}{3}\eta\, ,\label{zeta}
\end{align}
and $c_v$ is given by (\ref{cv}).
Here the integrals 
$D$ and $E$ are defined by
\begin{align}\nonumber
D\equiv D(\gamma) &\equiv \frac{1}{K_2(\gamma)}\int_0^\infty \gamma^{\star 2} K_1(\gamma^\star)\Phi(\mathcal{I})d\mathcal{I} \quad\hbox{and} \\\label{intCDE}
E\equiv E(\gamma) &\equiv \frac{1}{K_2(\gamma)}\int_0^\infty \gamma^{\star 2} \mathrm{Ki}_1(\gamma^\star)\Phi(\mathcal{I})d\mathcal{I} \, , 
\end{align}
respectively.

\subsection{Thermodynamic coefficients for polyatomic gases}
 Here we will consider the classical model, described in section \ref{preliminaries}, for the internal energies of  gases \cite{Pavic}, where the density of states are given by polytropes $\Phi({\cal{I}})\propto{\cal{I}}^\alpha$.
These models can give a fairly good description of the internal 
energies of polyatomic gases for not too high energies,  but since the adiabatic index $\alpha=(D-5)/2$ depends on the number of degrees of freedom, $D$, $\alpha$ is only constant in certain temperature intervals, where the number of degrees of freedom may be considered as constant. 
\vskip5mm

\subsubsection{Monoatomic gases}
For a monoatomic gas without internal degrees of freedom, the integrals $A$, $B$, $C$, $D$ and $E$ simplify to
\begin{equation}\label{ABCDEmono}
A=1\, , \quad B=\gamma\frac{K_3}{K_2}\, , \quad C=\gamma\, , \quad D=\gamma^2\frac{K_1}{K_2}\,\quad\hbox{and}\quad E=\gamma^2\frac{Ki_1}{K_2}\, .
\end{equation}
Hence the energy density and pressure are
\begin{equation}
\mu = nkT\left[\gamma\frac{K_3}{K_2}-1\right]\quad \hbox{and} \quad p=nkT
\end{equation}
and the corresponding thermodynamic coeffients, $\eta$, $\zeta$ and $\kappa$ are
\begin{equation}
\eta = \frac{nkT\tau\gamma}{15}\left(3\frac{K_3}{K_2}-\gamma+\gamma^2\frac{K_1}{K_2}-\gamma^2\frac{Ki_1}{K_2}\right)\, ,
\end{equation}
\begin{equation}
\zeta = -nkT\tau\frac{5\gamma\frac{K_3}{K_2}-\gamma^2\frac{K_3^2}{K_2^2}+\gamma^2}{5\gamma\frac{K_3}{K_2}-\gamma^2\frac{K_3^2}{K_2^2}+\gamma^2-1}+\frac{5}{3}\eta\, ,
\end{equation}
\begin{equation}
\kappa = \frac{nk^2T\tau}{3m}\gamma^4\frac{K_3}{K_2}\left(\frac{K_3}{K_2}\left(\frac{1}{\gamma}-\frac{K_1}{K_2}+\frac{Ki_1}{K_2}\right)-\frac{3}{\gamma^2}\right)\, ,
\end{equation}
which agree with equations (15) and (16) in \cite{Kremer1}. Their non-relativistic, $\gamma \rightarrow \infty$, and ultra-relativistic,
$\gamma \rightarrow 0$, limits are
\begin{equation}\nonumber
\mu =nmc^2+\frac{3}{2}nkT\left(1+\frac{5}{4\gamma}\right)\, , \quad
\kappa=\frac{5}{2m}nk^2T\tau\left(1-\frac{3}{\gamma}\right)\, ,
\end{equation}
\begin{equation}\label{monoNRlimit}
\eta=nkT\tau\left(1-\frac{1}{\gamma}\right)\, , \quad
\zeta=\frac{5}{6\gamma^2}nkT\tau\left(1-\frac{16}{\gamma}\right)\, 
\end{equation}
and
\begin{equation}\nonumber
\mu =3nkT\left(1+\frac{1}{6}\gamma^2\right) \, , \quad
\kappa=\frac{4}{3m}nk^2T\tau\gamma\left(1-\frac{11}{8}\gamma^2\right)\, ,
\end{equation}
\begin{equation}
\eta=\frac{4}{5}nkT\tau\left(1+\frac{1}{24}\gamma^2\right)\, , \quad
\zeta=\frac{1}{54}nkT\tau\gamma^4\left(1-\frac{3\pi}{2}\gamma\right)\, ,
\end{equation}
respectively.

\vskip5mm
\subsubsection{Diatomic gases}

A diatomic molecule with both rotational degrees of freedom excited has $D=5$. 
Hence the adiabatic index $\alpha$ equals zero, so that
$\Phi({\cal{I}})$ equals a nonzero constant with dimension of inverse energy. Since only ratios of the functions $A$, $B$, $C$, $D$ and $E$  appear in measurable quantities, we put this constant to one. For this choice they become\footnote{Note that there are some misprints in section 5 of \cite{ORS}.}
\begin{align}\nonumber
A(\gamma) &= \frac{mc^2}{\gamma}\frac{K_1(\gamma)}{K_2(\gamma)} \ , \\\nonumber
B(\gamma) &= \frac{mc^2}{\gamma}\left[2\frac{K_1(\gamma)}{K_2(\gamma)}+ \gamma\right] \ , \\\nonumber
C(\gamma) &=\frac{mc^2}{\gamma}\left(2+\frac{K_1(\gamma)}{K_2(\gamma)}\left(\gamma-\frac{4}{\gamma}\right)\right)\, , \\\nonumber
D(\gamma) &=mc^2\gamma
\ , \\\label{intABCDE2}
E(\gamma) &= \frac{mc^2\gamma}{3}\left[2+\gamma\left(\frac{K_1(\gamma)}{K_2(\gamma)}-\frac{Ki_1(\gamma)}{K_2(\gamma)}\right)\right]
\, ,
 \end{align}
respectively. 
The density and pressure are given by
\begin{equation}
\mu=nkT\left(1+\gamma\frac{K_2(\gamma)}{K_1(\gamma)}\right)\quad\hbox{and}\quad  p=nk T\, ,
\end{equation}
and the thermodynamical coefficients $\zeta$, $\eta$ and $\kappa$ by
\begin{eqnarray}\nonumber
\zeta &=& n k T\tau
\left[\frac{\left(1-\gamma^2\right)}{9}-\frac{1}{1+\frac{3\gamma K_2}{K_1}+\gamma^2\left(1-\frac{K_2^2}{K_1^2}\right)}+
+\frac{\gamma K_2}{9 K_1}\left(1+\frac{\gamma^2}{3}\right)\right.\\\nonumber
&&\left.+\frac{\gamma^4}{27}\left(\frac{Ki_1}{K_1}-1\right)
\right]\, ,\\\nonumber
\eta&=&\frac{\gamma n k T \tau}{15}\left[\frac{10}{\gamma}-\gamma-\frac{\gamma^3}{3}\left(1-\frac{Ki_1}{K_1}\right)+\left(1+\frac{\gamma^2}{3}\right)\frac{K_2}{K_1}\right]\, ,\\\nonumber
\kappa&=&\frac{\gamma n k^2 T \tau}{3m}\left(2+\gamma\frac{K_2}{K_1}\right)\left[\left(3-\gamma\frac{K_2}{K_1}+\gamma^2\left(1-\frac{Ki_1}{K_1}\right)\right)\left(2+\gamma\frac{K_2}{K_1}\right)\right.\\
&&\left.-3\right]\, .
\end{eqnarray}

In the non-relativistic limit $\gamma \rightarrow \infty$ these quantities go over into
\begin{equation}\nonumber
\mu=n m c^2+\frac{5}{2}n k T+\frac{3}{8\gamma}n k T\, , \quad
\kappa=\frac{7}{2m}n k^2 T\tau\left(1-\frac{37}{7\gamma}\right)\, ,
\end{equation}
\begin{equation}\label{DiatomicNR}
\eta=n k T\tau\left(1-\frac{1}{\gamma}\right)\, , \quad
\zeta=\frac{4}{15}n k T \tau\left(1-\frac{29}{5\gamma}\right)\, ,
\end{equation}
to leading orders in $1/\gamma$. For the asymptotic expansions of the Bickley-Naylor functions, see \cite{Milgram}. Note that the bulk viscosity vanishes in the non-relativistic limit for a monoatomic gas, (\ref{monoNRlimit}), in line with the result in \cite{Israel} for a simple gas, but not for a diatomic gas, (\ref{DiatomicNR}).

For the ultra-relativistic case, $\gamma\rightarrow 0$, the limits are
\begin{equation}\nonumber
\mu=3n k T\left(1-\frac{\ln\frac{\gamma}{2}+\tilde\gamma}{3}\gamma^2\right)\, ,
\end{equation}
\begin{equation}\nonumber
\kappa=\frac{4 c^2 n k\tau}{3}\left(1+\left(\frac{11}{4}\left(\ln\frac{\gamma}{2}+\tilde\gamma\right)+4\right)\gamma^2\right)\, ,
\end{equation}
\begin{equation}\nonumber
\eta=\frac{4}{5}n k T\tau\left(1-\frac{1}{12}\left(\ln\frac{\gamma}{2}+\frac{1}{3}+\tilde\gamma\right)\gamma^2\right)\, ,
\end{equation}
\begin{equation}
\zeta=\frac{2}{27}n k T\tau\gamma^2\left(1-\frac{1}{2}\left[\left(\ln\frac{\gamma}{2}+\tilde\gamma\right)\left(4\ln\frac{\gamma}{2}+4\tilde\gamma+3\right)+2\right]\right)\, ,
\end{equation}
where $\tilde\gamma=0.577 215 66...$ is the Euler-Mascheroni constant.

\section{Solutions to Einstein-Boltzmann's equations}\label{secLRS}

In this section we consider some homogeneous and Locally Rotationally Symmetric (LRS)  Bianchi VIII models with heat flow and vorticity, see, e.g. \cite{BradleySviestins}, 
as well as some orthogonal models without heat flow and vorticity. 
As source term in the field equations we use the energy-momentum tensor derived from the kinetic theory in section \ref{secThermocoeff}.

\subsection{Cosmological models of Bianchi type VIII with heat flow}\label{secBVIII}

The metric is given by the line element
\begin{equation}\label{BianchiVIIImetric}
ds^2=(cdt+a(t)\sigma^1)^2-b^2(t)(\sigma^1)^2-g^2(t)\left((\sigma^2)^2+(\sigma^3))^2\right)\, ,
\end{equation}
where the 1-forms $\sigma^i$ satisfy the Bianchi VIII structure relations
\begin{equation}
d\sigma^1=-\sigma^2\wedge\sigma^3\;,\quad d\sigma^2=\sigma^3\wedge\sigma^1;\ , \quad d\sigma^3=\sigma^1\wedge\sigma^2 \; .
\end{equation}
They can be given as
\begin{eqnarray}\nonumber
\sigma^1&=&dx+(1+x^2)dy+(x-y-x^2y)dz\, ,\\\nonumber
\sigma^2&=&2xdy+(1-2xy)dz\, ,\\
\sigma^3&=&dx+(-1+x^2)dy+(x+y-x^2y)dz\, ,
\end{eqnarray}
in the coordinates $x,y,z$.

Quantities will be given in a Lorentz frame comoving with the fluid (the particle frame). As form basis we choose
\begin{equation}
\omega^0=cdt+a(t)\sigma^1\; , \quad \omega^1=b(t)\sigma^1\; ,\quad \omega^2=g(t)\sigma^2\; ,\quad \omega^3=g(t)\sigma^3\, ,
\end{equation}
with the corresponding vector basis
\begin{eqnarray}\nonumber
X_0&=&\frac{\partial}{c\partial t}\, ,\\\nonumber
X_1&=&\frac{1}{b}\left[-a\frac{\partial}{c\partial t}+\frac{1}{2}(1+x^2)\frac{\partial}{\partial x}+\frac{1}{2}(1-2xy)\frac{\partial}{\partial y}-x\frac{\partial}{\partial z}\right]\, ,\\\nonumber
X_2&=&\frac{1}{g}\left[-x\frac{\partial}{\partial x}+y\frac{\partial}{\partial y}+\frac{\partial}{\partial z}\right]\, ,\\\label{tetrad}
X_3&=&\frac{1}{2g}\left[(1-x^2)\frac{\partial}{\partial x}-(1-2xy)\frac{\partial}{\partial y}+2x\frac{\partial}{\partial z}\right]\; .
\end{eqnarray}

Instead of the usual metric approch to the field equations, we will use  the 1+1+2 covariant split of spacetime, see \cite{Clarkson}, to describe the spacetime. It uses
covariant objects as dependent variables, which then are subjects to certain integrability conditions. With this method it will be easier to impose the results from the kinetic theory on the  different physical quantities.  We also obtain a first order system of differential equations, faciliating the check  of integrability. 
This form is also suitable for numerical integration. For reference, the Einstein tensor, Weyl tensor and Ricci rotation components in terms of the metric are given in appendix \ref{EinsteinWeylRicci}.

\subsubsection{New variables}

Due to the LRS symmetry all necessary variables become scalars in the 1+1+2 formalism, \cite{Clarkson}. These can be chosen among the components of the Einstein and Weyl tensors, as well as 
the kinematic quantites of the fluid velocity $u=cX_0$ ($u^i=c\delta^i_0$) and of the vector $n^i=\delta^i_1$ along the symmetry axis.

The nonzero kinematic quantities of $u^i$, corresponding to the Ricci rotation coefficents $\gamma^0_{\; ij}$ (see (\ref{kingam})), are
\begin{eqnarray}\nonumber
\theta&\equiv& c\bar\theta\, , \quad  a_1 = a_i n^i\equiv -c^2{\cal{A}}\, , \quad  \omega_{23}=-\frac{1}{2}\varepsilon^{ijk}n_i \omega_{jk}\equiv c\Omega\, , \\
 \sigma_{11}&=&-2\sigma_{22}=-2\sigma_{33}=n^i n^j \sigma_{ij}\equiv -c\Sigma \, ,
\end{eqnarray}
where  $\varepsilon _{ijk}\equiv\varepsilon_{ijkl} \frac{u^l}{c}$, and $\varepsilon_{ijkl}$ is the completely antisymmetric object with $\epsilon_{0123}=+1$.
We have here introduced the scalars $\bar\theta$, ${\cal{A}}$, $\Omega$ and $\Sigma$, to make a comparison with \cite{Clarkson} easier. That paper uses the conventions $c=8\pi G=1$ and has the signature $+2$. Consequently we also define
\begin{equation}
\bar{t}\equiv ct \, ,
\end{equation}
and hence an overdot will indicate derivative with respect to $\bar{t}$.

Two more scalars, the twist, $\xi$,  and expansion, $\phi$, of the two-sheets perpendicular to the symmetry-axis $n^i=\delta_1^i$, corresponding to the Ricci rotation coefficients $\gamma^1_{\; ab}$,
are given by
\begin{equation}\label{xiphi}
\xi\equiv\frac{1}{2}\varepsilon^{ijk}n_k n_{i;j}
\quad\hbox{and}\quad \phi\equiv -h^{ij}n_{i;j}
\, .
\end{equation}

Einstein's equations are imposed by expressing the Einstein tensor in terms of the energy-momentum tensor and the cosmological constant $\Lambda$
\begin{eqnarray}\nonumber
G_{00}&=&\frac{8\pi G}{c^4}T_{00}=\frac{8\pi G}{c^4}\mu+\Lambda\equiv \bar\mu+\Lambda\\\nonumber
G_{01}&=&\frac{8\pi G}{c^4}T_{01}=\frac{8\pi G}{c^4}\frac{q_1}{c}\equiv -\tilde{Q}\\\nonumber
G_{11}&=&\frac{8\pi G}{c^4}T_{11}=\frac{8\pi G}{c^4}\left(p+\Pi+\pi_{11}\right)-\Lambda\equiv \bar{p}+\bar{\Pi}-{\tilde{\Pi}}-\Lambda\\\nonumber
G_{22}&=&G_{33}=\frac{8\pi G}{c^4}T_{22}=\frac{8\pi G}{c^4}T_{33}=\frac{8\pi G}{c^4}\left(p+\Pi-\frac{1}{2}\pi_{11}\right)-\Lambda\\\label{EE}
&\equiv&\bar{p}+\bar{\Pi}+\frac{1}{2}{\tilde{\Pi}}-\Lambda \, .
\end{eqnarray}
Similarly to above, we have here also introduced the scalars $\bar\mu$, $\bar p$, $\tilde Q$, $\bar\Pi$ and $\tilde\Pi$ 
\begin{eqnarray}\nonumber
\bar\mu&\equiv& \frac{8\pi G}{c^4}\mu\, , \quad \bar{p}\equiv \frac{8\pi G}{c^4}p\, , \quad \bar\Pi\equiv \frac{8\pi G}{c^4}\Pi\,\\
\tilde{Q}&\equiv& -\frac{8\pi G}{c^5}n^i q_i =-\frac{8\pi G}{c^5}q_1\, , \quad \tilde\Pi\equiv -\frac{8\pi G}{c^4}n^i n^j \pi_{ij}=-\frac{8\pi G}{c^4}\pi_{11}
\end{eqnarray}
to fit the notation of \cite{Clarkson}.

The remaining parts of the Riemann tensor are given in terms of the electric and magnetic parts of the Weyl tensor
\begin{equation}
E_{ij}\equiv C_{ijkl}\frac{u^j}{c}  \frac{u^l}{c} \quad \hbox{and} \quad H_{ij}\equiv\frac{1}{2}\varepsilon_{ilm} C^{lm}_{\;\;\; jk} \frac{u^k}{c},
\end{equation}
whose nonzero components are
 \begin{eqnarray}\label{WeylE}\nonumber
 E_{11}&=&-2E_{22}=-2E_{33}=n^i n^j E_{ij}\equiv -{\cal{E}}
 \end{eqnarray}
 and
 \begin{equation}\label{WeylH}
 H_{11}=-2H_{22}=-2H_{33}=n^i n^j H_{ij}\equiv -{\cal{H}}
 \end{equation}
 respectively. Hence the components are completely determined by the two scalars ${\cal{E}}$ and ${\cal{H}}$. %
 
 Our variables, corresponding to the 1+1+2 scalars in the covariant 1+1+2 split of spacetime \cite{Clarkson}, are hence given by the set
\begin{equation}\label{setS}
S=\{{\cal{E}},{\cal{H}},\bar\mu,\bar{p},\bar\Pi,{\tilde\Pi},\tilde{Q},{\cal{A}},\bar\theta,\Sigma,\Omega,\phi,\xi\}\, ,
\end{equation}
where the scalars ${\cal{E}}$ and ${\cal{H}}$ fully describe the electric, $E_{ij}$, and magnetic,$ H_{ij}$, parts of the Weyl tensor
, $\bar\mu$ is the energy density, $\bar{p}$ is the
isotropic pressure, $\bar\Pi$ is the bulk pressure, ${\tilde\Pi}$ determines the anisotropic pressure and $\tilde{Q}$ is the only nonzero component of the heat flow. 
The kinematic quantites acceleration, expansion, shear and vorticity are completely given in terms of the scalars ${\cal{A}}$, $\bar\theta$, $\Sigma$, and $\Omega$, respectively.
Finally, $\phi$ and $\xi$, correspond to the expansion and twist, respectively, of the 2-sheets perpendicular to the symmetry axis $n^i=\delta^i_1$.

\subsubsection{Evolution equations}
 
From the set $S$ one may, together with how the directional derivatives, $X_i$, act, construct a complete local description of the spacetime. See, e.g. \cite{BradleyMarklund}, for how such a description can be given in terms of the Riemann tensor and a finite number of its covariant derivatives in a fixed frame, or equivalently by the Riemann tensor, part of the Ricci rotation coefficents and part of the directional derivatives. For a set $S$ to describe a spacetime certain integrability conditions have to be imposed. In the present case sufficient conditions are the Ricci identies for $u^i$ and $n^i$, the Bianchi identities and the commutator relations.
 
Since spacetime is homogeneous, all of the quantities are functions of time $\bar{t}$ only. Any combination of them would do as the essential coordinate, so we keep $\bar{t}$ as coordinate.
The nonzero directional derivatives are hence given by 
\begin{equation}\label{derivatives}
X_0(S)=\frac{\partial S}{\partial {\bar{t}}}\quad\hbox{and}\quad X_1(S)=-\frac{a}{b}\frac{\partial S}{\partial {\bar{t}}}
\end{equation}
respectively. The factor $a/b$ describes the tilt of the normal of the hypersurfaces of homogeneity to the 4-velocity of matter. %
From the commutator equation 
\begin{equation}\label{Commutator}
\left[X_i, X_j\right]S+2\gamma^k_{\; [ij]}X_k S=0 \, ,
\end{equation}
(with $ij=23$), it follows that $a/b=-\Omega/\xi$, i.e. it is determined by the elements in $S$.

On imposing the Ricci identities for the 4-velocity, $u^i$, and the unit vector along the symmetries axis, $n^i$,
\begin{equation}
u_{i;jk}-u_{i;kj}=R^m_{\; ijk}u_m\, ,\quad \hbox{and} \quad n_{i;jk}-n_{i;kj}=R^m_{\; ijk}n_m\, ,
\end{equation}
respectively, the Bianchi identities
\begin{equation}
R^i_{\;\; jkl;m}+R^i_{\;\; jmk;l}+R^i_{\;\; jlm;k}=0\, ,
\end{equation}
and the commutator relations (\ref{Commutator}),
a system of first order differential equations and algebraic constraints is obtained. This system agrees with what one obtains from the formalism in \cite{Clarkson}, when applied to homogeneous LRS spacetimes. 

By successive elimination from the constraints, and differentiation of the constraints, the system will now reduce to the following algebraic equations
\begin{eqnarray}\nonumber
\phi&=&-\left(\Sigma-\frac{2}{3}\bar\theta\right)\frac{\Omega}{\xi}\\\nonumber
\tilde{Q}&=&-\frac{\Omega\xi}{\Omega^2+\xi^2}\left(\bar\mu+\bar{p}+\bar\Pi+\tilde\Pi\right)\\\nonumber
{\cal{H}}&=&\frac{1}{3\xi}\left[9\xi^2\Sigma-6\Omega\xi{\cal{A}}-3\left(\Sigma-\frac{2}{3}\bar\theta\right)\Omega^2\right]\\\nonumber
{\cal{E}}&=&\frac{\Omega}{3}\left(\Sigma-\frac{2}{3}\bar\theta\right)+\frac{1}{3}\left(\bar\mu+3\bar{p}+3\bar\Pi+\frac{3}{2}\tilde\Pi-2\Lambda\right)-2\left(\Omega^2-\xi^2\right)\\\label{algeq}
&&+\frac{2}{9}\left(\bar\theta^2-\frac{9}{2}\Sigma^2+\frac{3}{2}\Sigma\bar\theta\right)-\frac{\xi^2}{\Omega^2+\xi^2}\left(\bar\mu+\bar{p}+\bar\Pi+\tilde\Pi\right)
\end{eqnarray}
and the evolution equations
\begin{eqnarray}\label{xieq}%
\dot\xi&=&\frac{\xi}{3}\left(6\Sigma-\bar\theta\right)\\\label{Omegaeq}
\dot\Omega&=&\left(\Sigma-\frac{2}{3}\bar\theta\right)\Omega+{\cal{A}}\xi\\\nonumber
\dot{\bar\theta}-\frac{\Omega}{\xi}\dot{\cal{A}}&=&{\cal{A}}\left({\cal{A}}-\frac{\Omega}{\xi}\left(\Sigma-\frac{2}{3}\bar\theta\right)\right)-\frac{1}{3}\bar\theta^2-\frac{3}{2}\Sigma^2+2\Omega^2
\\\label{thetaeq}
&&-\frac{1}{2}\bar\mu-\frac{3}{2}\bar{p}-\frac{3}{2}\bar\Pi+\Lambda\\\nonumber
\dot\Sigma-\frac{2\Omega}{3\xi}\dot{{\cal{A}}}&=&\frac{2}{3}{\cal{A}}\left({\cal{A}}-\frac{\Omega}{\xi}\left(\Sigma-\frac{2}{3}\bar\theta\right)\right)-\Sigma\bar\theta+\frac{1}{2}\Sigma^2
-\frac{2}{9}\bar\theta^2+\frac{2}{3}\Lambda-\\\label{Sigmaeq}
&&2\xi^2+\frac{4}{3}\Omega^2+\frac{2}{3}\bar\mu+\tilde\Pi-\frac{\Omega^2}{\Omega^2+\xi^2}\left(\bar\mu+\bar{p}+\bar\Pi+\tilde\Pi\right)\\\nonumber
\end{eqnarray}
\begin{eqnarray}\nonumber
\dot{\bar\mu}-\frac{\Omega^2}{\xi^2}\left(\dot{\bar{p}}+\dot{\bar\Pi}+\dot{\bar{{\tilde{\Pi}}}}\right)=\frac{\Omega\left(\Omega^2+3\xi^2\right)}{\xi\left(\Omega^2+\xi^2\right)}{\cal{A}}\left(\bar\mu+\bar{p}+\bar\Pi+\tilde\Pi\right)-\frac{7\Omega^2+3\xi^2}{2\xi^2}\Sigma\tilde\Pi\\\label{mueq}
+\frac{\Omega^2}{3\xi^2}\tilde\Pi\bar\theta+
\frac{2\Omega^4}{3\xi^2\left(\Omega^2+\xi^2\right)}\tilde\Pi\left(3\Sigma+\bar\theta\right)
-\frac{\Omega^2\left(2\Sigma+\frac{5}{3}\bar\theta\right)+\xi^2\bar\theta}{\Omega^2+\xi^2}\left(\bar\mu+\bar{p}+\bar\Pi\right)\, .
\end{eqnarray}

\subsubsection{In terms of kinetic theory}

In line with the conventions in \cite{Clarkson} and the definitions of the quantities in the set (\ref{setS}), we also rewrite the parameters of the kinetic theory: relaxation time, $\tau$, temperature, $T$, particle mass $m$, specific heat, $c_v$ and the coefficients of bulk viscosity, $\zeta$, shear viscosity, $\eta$, and conductivity, $\kappa$, in terms of $\bar\tau$, 
$\bar{T}$, $\bar{m}$, $\bar{c}_v$, $\bar\zeta$, $\bar\eta$ and $\bar\kappa$, respectively as
\begin{eqnarray}\nonumber
\tau&=&\bar\tau/c\, , \quad T=\frac{c^4}{8\pi G k}\bar{T}\, ,\quad m=\frac{c^2}{8\pi G}\bar{m}\, , \quad c_v=k\bar{c}_v\, , \\\label{unit2}
 \zeta&=&\frac{c^3}{8\pi G}\bar\zeta\, ,\quad \eta=\frac{c^3}{8\pi G}\bar\eta\, ,\quad \kappa=ck\bar\kappa\, .
\end{eqnarray}

From the distribution function, we have that the variables which determine the energy-momentum tensor, $\bar\mu$, $\bar{p}$, $\bar\Pi$, $\tilde\Pi$ and $\tilde Q$,
are all functions of $n$ and $\bar{T}$ and the kinematic quantities of the particle velocity $u^i$.
In terms of the quantities (\ref{setS}) and the  definitions (\ref{unit2}), equations
(\ref{mupeq}), (\ref{eq:Pieq}) and (\ref{eq:pieq})  can be rewritten as
\begin{equation}\label{kineq}
\bar\mu=n\bar{T}\left[\frac{B}{A}-1\right]\, ,\quad \bar{p}=n\bar{T}\, ,\quad \bar\Pi=-\bar\zeta\bar\theta\, ,\quad \tilde\Pi=-2\bar\eta\Sigma\, .
\end{equation}
Similarly we can rewrite the heat conductivity equation (\ref{eq:qeq}) as
\begin{equation}\label{Qeq}
\tilde{Q}=-\bar\kappa\left(\frac{\Omega}{\xi}\dot{\bar{T}}+{\cal{A}}\bar{T}\right)\, .
\end{equation}
Substitution of (\ref{Qeq}) into (\ref{algeq}b) gives
an evolution equation for the temperature
\begin{equation}\label{timeeq}
\dot{\bar{T}}=\frac{\xi^2}{\bar\kappa\left(\Omega^2+\xi^2\right)}\left(\bar\mu+\bar{p}+\bar\Pi+\tilde\Pi\right)-\frac{\xi}{\Omega}{\cal{A}}\bar{T} \, .
\end{equation}
To close the system, we also add the time evolution equation for the particle density (\ref{ndens})
\begin{equation}\label{parteq}
\dot{n}=-n\bar\theta \, .
\end{equation}

Substitution of (\ref{kineq}) into equation (\ref{mueq}) gives an equation involving the time derivatives of $n$, $\bar{T}$, $\bar\theta$ and $\Sigma$. Substitution
of equations (\ref{thetaeq}),({\ref{Sigmaeq}) 
and (\ref{parteq}) into (\ref{mueq}) then gives the following time evoultion equation for the acceleration 
${\cal{A}}$
\begin{align}
\dot{\mathcal{A}} = &~ -\mathcal{A}\left(\frac{\xi}{\Omega}\mathcal{A} -\left(\Sigma - \frac{2\bar\Theta}{3}\right)\right)
-\frac{3\bar{\tilde{\Pi}}}{2\tilde{D}}\frac{\xi^3}{\Omega^3}\left(\Sigma-\frac{2\bar\Theta}{3}\right)   
\notag \\
    &+\frac{\xi}{\Omega}\Bigg[\frac{\bar\Theta^2}{3}+\frac{3\Sigma^2}{2}-2\Omega^2+\frac{\bar\mu}{2}+\frac{3\bar{p}}{2}+ \frac{3\bar\Pi}{2}-\Lambda +\frac{4\bar\eta\xi^2}{\tilde{D}} \notag \\
    &+ \frac{\bar\mu}{\tilde{D}}\left(1+\frac{\xi^2}{\Omega^2}\right)\bar\Theta + \left(n-\frac{\xi^2}{\Omega^2}\pdv{\bar\mu}{\bar{T}}-\pdv{\bar\zeta}{\bar{T}}\bar\Theta-2\pdv{\bar\eta}{\bar{T}}\Sigma\right)\frac{\dot{\bar{T}}}{\tilde{D}} \Bigg] \notag  \\
    &+\frac{\tilde{Q}}{\tilde{D}}\left(\frac{2\xi^2}{\Omega^2}\left(\Sigma+ \frac{4\bar\Theta}{3}+\bar\eta\right) + \bar\Theta\left(1+\frac{\xi^4}{\Omega^4}\right) - \frac{\xi}{\Omega}\left(1+\frac{3\xi^2}{\Omega^2}\right)\mathcal{A} \right), \label{eqAdot}
\end{align}
where
\begin{align}\label{eqDD}
\tilde{D} = &~\bar\zeta + \frac{4\bar\eta}{3}\, , 
\end{align}
and $\tilde{Q}$ and $\dot{\bar{T}}$ are given by (\ref{algeq}b)  and (\ref{timeeq}) respectively.
Here it has been used that $\bar\mu$, $\bar{p}$, $\bar\zeta$ and $\bar\eta$ are functions of $n$ and $\bar{T}$, and that they are linear in $n$, so that
\begin{equation}
\frac{\partial\bar\mu}{\partial n}=\frac{\bar\mu}{n}\, ,\quad \frac{\partial\bar{p}}{\partial n}=\frac{\bar{p}}{n}=\bar{T}\, ,\quad \frac{\partial\bar\zeta}{\partial n}=\frac{\bar\zeta}{n}
\, , \quad \frac{\partial\bar\eta}{\partial n}=\frac{\bar\eta}{n}\, .
\end{equation}
Also
\begin{equation}
\frac{\partial\bar{p}}{\partial\bar{T}}=n\, .
\end{equation}
For the expressions of the kinematic coefficients and their derivatives in terms of the new variables, see appendix \ref{AppendixCoefficient}.
\vskip1cm
In summary we get a closed system of first order ordinary differential equations for the 7 variables $\xi$, $\Omega$, $\bar\theta$, $\Sigma$, ${\cal{A}}$, $\bar{T}$ and $n$
(equations (\ref{xieq}), (\ref{Omegaeq}), (\ref{thetaeq}), (\ref{Sigmaeq}), (\ref{eqAdot}), ({\ref{timeeq}) and (\ref{parteq})), 
wheras $\bar\mu$, $\bar p$, $\bar\Pi$ and $\tilde\Pi$ are given algebraically by (\ref{kineq}), and $\phi$, $\tilde{Q}$, ${\cal{E}}$ and ${\cal{H}}$ algebraically by (\ref{algeq}). 
From the solutions to this system, one may then obtain the scale factors $a$, $b$ and $g$ by integration of, e.g. $\gamma^0_{\;10}$, $\gamma^0_{\; 11}$ and  $\gamma^0_{\; 22}$ in (\ref{RicciRot}). Since the conditons for integrability already are imposed in the above procedure, the remaining equations (\ref{RicciRot}) and the equations (\ref{EE}), (\ref{WeylE}) and (\ref{WeylH}) will all be automatically satisfied.

\subsection{The orthogonal case}
The orthogonal case is obtained by choosing the tilt factor $a(t)=0$ in the metric (\ref{BianchiVIIImetric}). The varibles $\phi$, $\tilde{Q}$, $\Omega$ and ${\cal{A}}$ then all become zero,
and the system reduces to
\begin{eqnarray}\nonumber
\dot\xi&=&\frac{\xi}{3}\left(6\Sigma-\bar\theta\right)\\\nonumber
\dot{\bar\theta}&=&-\frac{1}{3}\bar\theta^2-\frac{3}{2}\Sigma^2
-\frac{1}{2}\bar\mu-\frac{3}{2}\bar{p}-\frac{3}{2}\bar\Pi+\Lambda\\\nonumber
\dot\Sigma&=&-\Sigma\bar\theta+\frac{1}{2}\Sigma^2
-\frac{2}{9}\bar\theta^2+\frac{2}{3}\Lambda+\frac{4}{3}\Omega^2+\frac{2}{3}\bar\mu+\tilde\Pi-2\xi^2\\\nonumber
n\bar{c}_v\dot{\bar{T}}&=&-\frac{3}{2}\Sigma\tilde\Pi
-\left(\bar{p}+\bar\Pi\right)\bar\theta\\\label{orthogon}
\dot{n}&=&-n\bar\theta\, ,
\end{eqnarray}
with ${\cal{H}}$ and ${\cal{E}}$ defined algebraically by
\begin{eqnarray}\nonumber
{\cal{H}}&=&3\xi\Sigma\\
{\cal{E}}&=&-\frac{2}{3}\bar\mu-\frac{2}{3}\Lambda-\frac{1}{2}\tilde\Pi+2\xi^2+\frac{2}{9}\bar\theta^2 -\Sigma^2+\frac{1}{3}\Sigma\bar\theta\, ,
\end{eqnarray}
and, as before, with $\bar\mu$, $\bar{p}$, $\bar\Pi$, $\bar\tilde\Pi$ and  $\bar{c}_v$ given by (\ref{kineq}) and  (\ref{cv}), respectively. The evolution equation
for the temperature, $\bar{T}$, now is obtained from the general equation (\ref{Ttimeder}).

For the system (\ref{orthogon}) to be of Bianchi class VIII, it is necessary for $\xi$ to be nonzero, but with $\xi\equiv 0$ the system actually describes the Kantowski-Sachs universa
\begin{equation}
ds^2=c^2dt^2-a_1^2(t)dz^2-a_2^2(t)\left(d\theta^2+\sin^2\theta d\varphi^2\right)\, ,
\end{equation} 
with
\begin{equation}
\theta=\frac{\dot{a}_1}{a_1}+2\frac{\dot{a}_2}{a_2}\quad \hbox{and}\quad\Sigma=\frac{2}{3}\left(\frac{\dot{a}_1}{a_1}-\frac{\dot{a}_2}{a_2}\right)\, .
\end{equation}

\section{Numerical solutions}\label{secNumerical}

Through the constitutive relations provided by the kinetic description, we see that the systems close and should, in principle, be solvable. However, as their lengthy nature hinders analytical solutions, we choose to investigate the dissipative Bianchi VIII equations numerically. For this, we consider the simplest non-trivial model including internal degrees of freedom, namely the polytropic case with adiabatic index $\alpha = 0$ so that $\Phi(\mathcal{I})$ is constant.  As described previously, this density of states could in certain temperature intervals be seen as describing a diatomic gas. However, in this section, we will view this model simply as a non-trivial example with internal degrees of freedom, rather than as modelling any specfic gas. }  

As will be seen below, the tilted Bianchi VIII models behave very differently than the orthogonal models with respect to the dissipative effects, and there is no natural limit relating the two cases. Rather, an intial tilt grows, eventually making the surfaces of homogeneity light-like or worse. This phenomena has also been observed in \cite{Hervik,Shogin}. A consequence of this is that high temperatures seem to be needed to maintain the required heat flow in tilted models, whereas this is not the case for orthogonal models. Hence the simulations for tilted models  apply to early and hot universa, whereas those for orthogonal models can describe also later universa close to thermodynamical equilibrium.  As such, we will treat the tilted and orthogonal models separately. 

After our rescaling to quantities with bars and tildes, the dimensions of all variables used are in powers of length, which means that there is only one scale $L$ left. In practice, we will implicitly assume that each variable has been further rescaled with a suitable power of $L$ to make them dimensionless. This allows us to treat $L$ as arbitrary for now. To get correct magnitudes for a particular length scale would then imply restoring $L$ by multiplying the dimensionless values with the corresponding powers of $L$.   What is ultimately important in the numerical simulations is therefore the relation between the variables rather than their particular values. In what follows, we will choose the absolute values of the initial conditions to be centered around unity, but this choice is ultimately arbitrary. More importantly, to facilitate numerical stability, particularly in the tilted case, we choose initial conditions that do not differ amongst each other by too many orders of magnitude. It should then be kept in mind that the specific solutions are not intended as models of our Universe, but rather as showing the self-consistency and solvability of the system. Finally, we drop the bars over rescaled quantities.

\subsection{The tilted case}
Starting with the tilted Bianchi VIII models, choosing the initial conditions and parameters as 
\begin{align} \label{eq:TilltedICs}
\Theta(0) &= 0.1,\quad \Sigma(0) = 0.001,\quad \xi(0) = 1, \quad \mathcal{A}(0) = 10, \quad \Omega(0) = 0.1,  \notag  \\
 \Lambda &= 1, \quad m = 1, \quad n(0) = 1, \quad T(0) = 1, \quad \tau =0.1 
\end{align} 
we obtain the results in Fig.~\ref{fig:IC1}. 

\begin{figure}[htb]
\makebox[\textwidth][c]{\includegraphics[width=\textwidth]{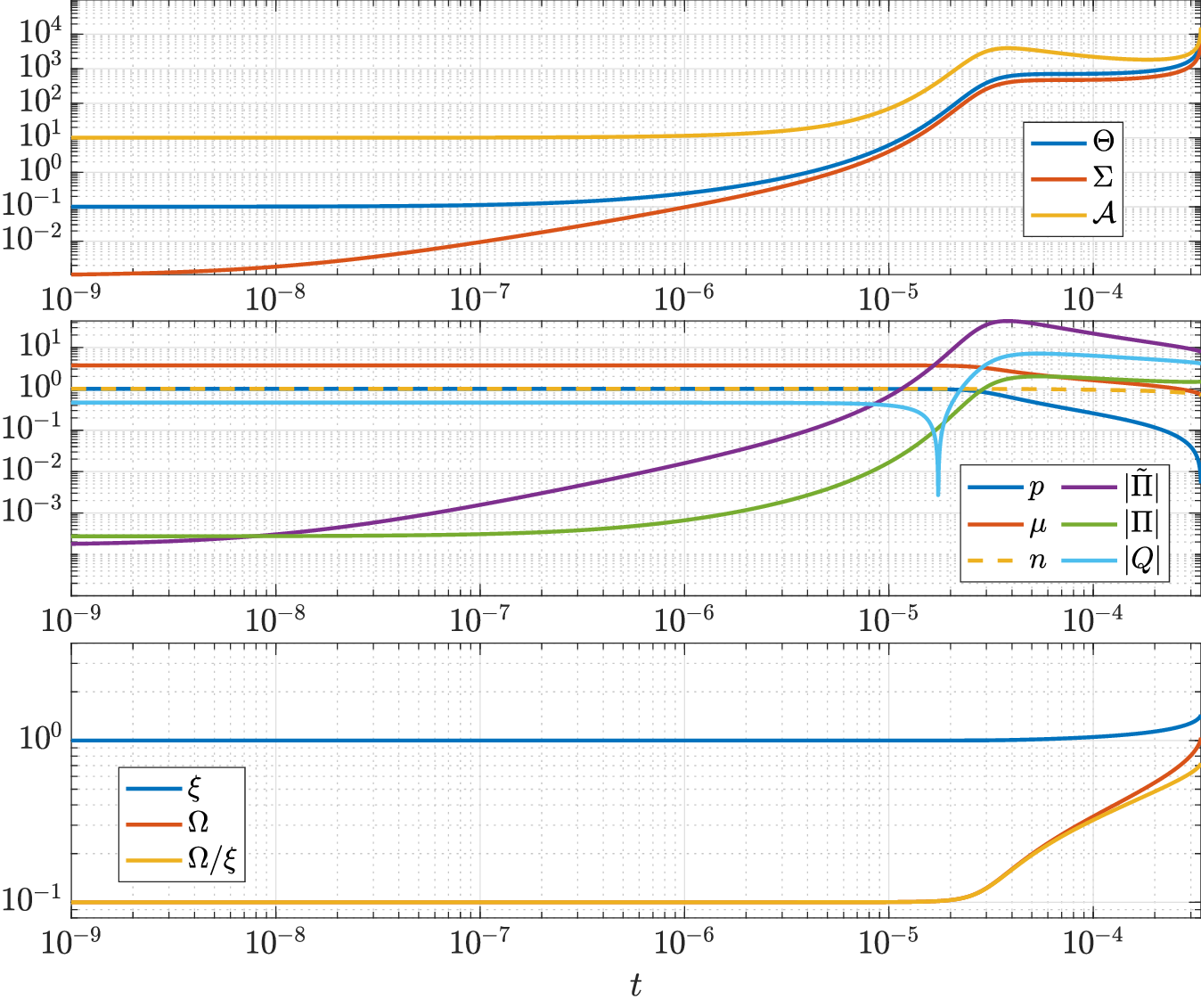}}
\caption{Numerical solution of the tilted Bianchi VIII system with initial conditions and parameters given by (\ref{eq:TilltedICs}).}
\label{fig:IC1}
\end{figure}

There it can be seen that the shear gradually grows to match the size of the expansion rate, after which both quantities enter a very rapid growing phase around $t\sim10^{-5}$ together with the acceleration. Then, after slowing down for a while, the growth rate appears to increase again, but before the quantities have had time to change significantly, the solution is terminated as the integrator is unable to meet the required tolerances without decreasing the step size below the smallest value allowed. 

The rapid increase in magnitude of $\Sigma$, $\Theta$, and $\mathcal{A}$, eventually leading to integrator failure, is found to be a typical scenario for the tilted system. However, at the onset of the very rapid growing phase in Fig.~\ref{fig:IC1}, it should be noted that the dissipative effects, in particular the viscous shear stress, have already become comparable to the equilibrium quantities, signaling a breakdown of the validity of the Chapman-Enskog near-equilibrium approach. The eventual violation of the near-equilibrium assumption appears to be a generic feature of the tilted models, but the time at which this breakdown occurs is found to depend on the initial conditions. To illustrate this dependence, we rescale the parameters in (\ref{eq:TilltedICs}) one at a time, except $m$ which is kept at unity. Defining the breakdown time as the time at which the magnitude of either the bulk or shear viscous stress becomes equal to the equilibrium pressure, we can then plot this time with respect to the size of the rescaling factor. This is shown in Fig.~\ref{fig:breakdown_IC1}, where we have also included the case when both $\Omega$ and $\xi$ are rescaled simultaneously, corresponding to a fixed initial tilt factor. 

\begin{figure}[htb]
\makebox[\textwidth][c]{\includegraphics[width=\textwidth]{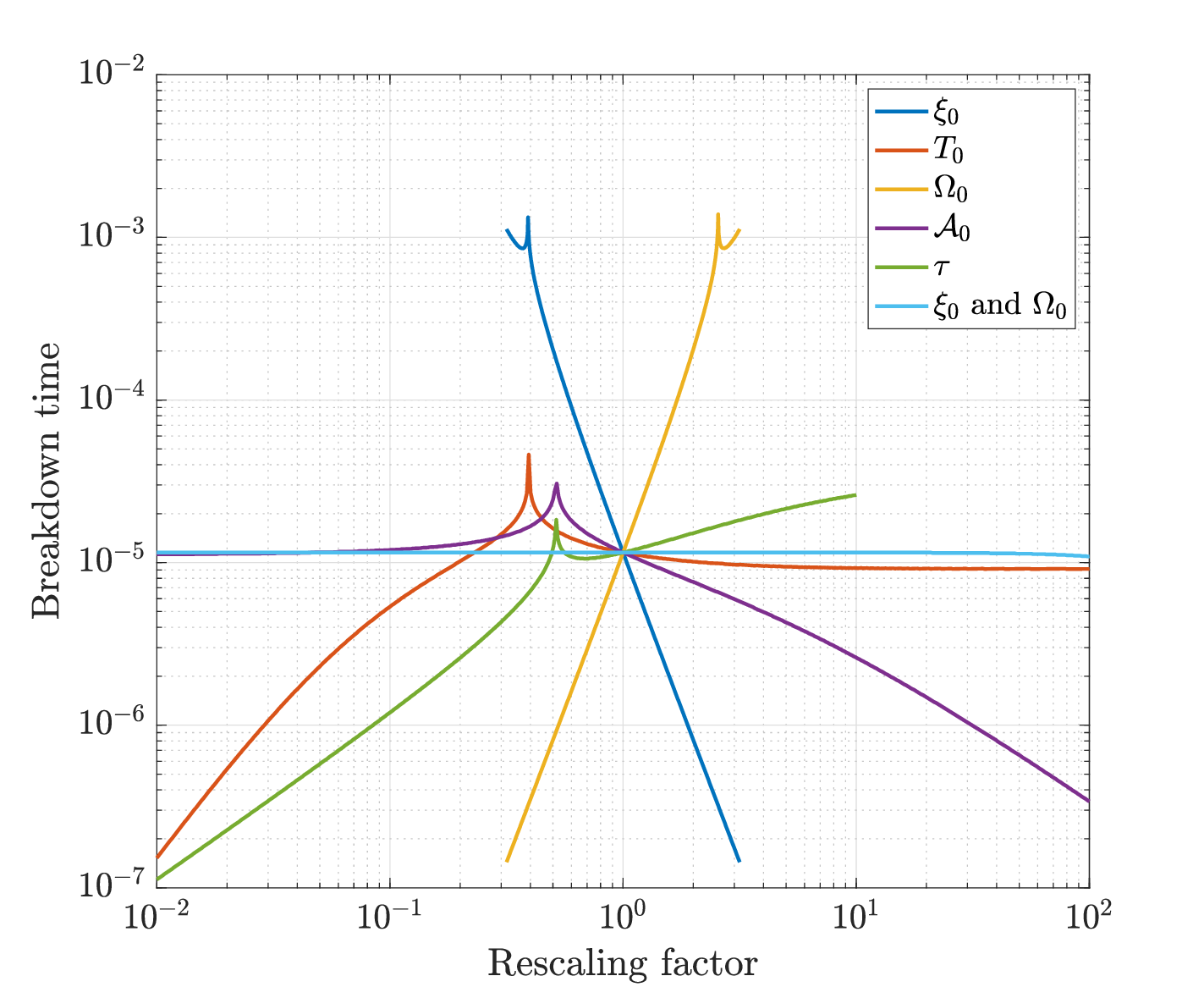}}
\caption{The breakdown time at which the dissipative stresses become comparable to the equilibrium pressure, for different initial conditions and parameters. The initial conditions and parameters are those in (\ref{eq:TilltedICs}), but with one (or two) multiplied with a rescaling factor. The breakdown time is plotted with respect to the size of this rescaling factor. The quantity that is being rescaled is indicated in the legend.}
\label{fig:breakdown_IC1}
\end{figure}

In Fig.~\ref{fig:breakdown_IC1}, we have only included curves for the rescaling of $\tau$, $\mathcal{A}$, $\Omega$, $\xi$, and $T$, as these are found to affect the breakdown time the most. Rescaling the remaining quantities does not change the order of magnitude of the breakdown time. Of the curves in Fig.~\ref{fig:breakdown_IC1}, the largest effects are obtained when rescaling $\Omega$ and $\xi$.  However, those effects can not be attributed to $\Omega$ and $\xi$ separately, as rescaling them at the same time barely affects the breakdown time, which is also illustrated in Fig.~\ref{fig:breakdown_IC1}. Thus, the important quantity is rather their ratio, i.e. the tilt factor. Increasing the tilt factor would therefore be an effective way to obtain a more stable model. However, to avoid the apparent instability, we find that the tilt factor has to be chosen greater than unity, which is impossible without allowing $u^a$ to become space-like.  

Another interesting feature in Fig.~\ref{fig:breakdown_IC1} is that the breakdown time decreases when decreasing the relaxation time $\tau$. As the dissipative Eckart theory is known to be plagued by instabilities due to its acausal nature, it would be natural to assume that the model should become stable if we decrease $\tau$ towards zero, approaching a perfect fluid case. However, this assumption is wrong due to the construction of the model. Since the tilt of the model requires a non-zero heat flow through (\ref{algeq}b), there is no natural limit of a tilted perfect fluid as $\tau$ decreases. In fact, there are no tilted LRS models with a perfect fluid as its only source \cite{King}.  Additionally, due to the divisions with $\Omega$ in the evolution equations for the tilted system, there is not any natural limit tending towards an orthogonal system either. As a result, instead of tending towards an orthogonal perfect fluid when decreasing $\tau$, what we observe is that, although the dissipative coefficients become smaller initially, the drivers of the dissipative stresses, e.g. the expansion rate and the shear, rapidly increase to compensate for that. A similar effect is seen when decreasing $T$.

As a final remark on Fig.~\ref{fig:breakdown_IC1}, we note that the cusp-like structures in the curves for $\xi_0$, $T_0$, $\Omega_0$, $\mathcal{A}_0$, and $\tau$ mark a transition between contracting and expanding final states. The final state is contracting for values of $\xi_0$, $T_0$, $\mathcal{A}_0$, and $\tau$ below their respective cusps, and expanding for values above the cusps. This situation is reversed for $\Omega_0$, which has expanding states below its cusp and contracting states above. Very near the cusps it may be difficult to assign the solution as being contracting or expanding, as it may oscillate back and forth between expansion and contraction.

\subsection{The orthogonal case}\label{secNumOrthogonal}
While the tilted models were found to be greatly affected by the dissipative effects and the precise value of $\tau$, this is not a property of the orthogonal system. In the orthogonal case, no divisions with $\tau$ appear in the evolution equations, and it is thus possible to obtain a perfect fluid by setting $\tau$ to zero. Comparing some perfect fluid solutions with their counterparts with non-zero $\tau$, we get the results in Fig.~\ref{fig:orthogonal_1}. As an illustrative example, we only plot the shear $\Sigma$. A complete plot of the solutions for the remaining variables can be found in appendix \ref{secPlot}. 
\begin{figure}[htb!]
\makebox[\textwidth][c]{\includegraphics[width=1.5\textwidth]{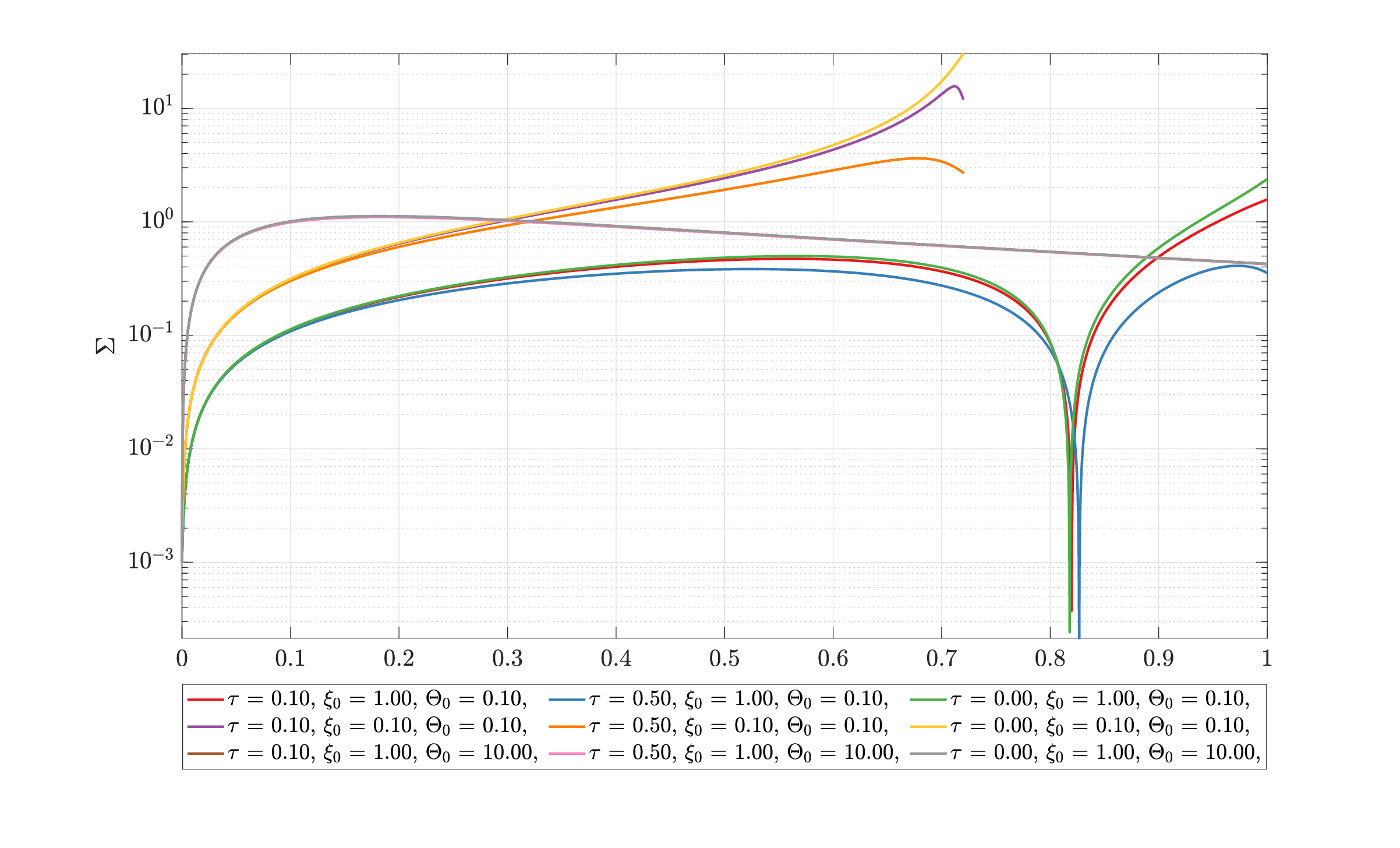}}
\caption{Solution for the shear $\Sigma$ in the orthogonal system for different initial conditions and parameters based on (\ref{eq:TilltedICs}), but with $\mathcal{A} = \Omega = 0$.  The initial conditions and parameters that differ between the curves are indicated in the legend.} 
\label{fig:orthogonal_1}
\end{figure}

In Fig.~\ref{fig:orthogonal_1}, it can be seen that solutions that differ only in their value of $\tau$, which share rows in the legend, are generally very similar. For the bottom row in the legend, they are so similar that they appear to coincide in the figure.  The largest difference can be found for the solutions that end in a contracting state, corresponding to the first two rows in the legend of Fig.~\ref{fig:orthogonal_1}. For the contracting solution with $\xi\left(0\right) = 1$, we note some visible differences in the evolution of $\Sigma$ but not in the other quantities (except $\Pi$ and $\tilde\Pi$ which are suitably rescaled by $\tau$, see appendix \ref{secPlot}). When  $\xi\left(0\right) = 0.1$, large differences can be clearly observed for the different values of $\tau$. However, when those differences start to become noticeable in the figure, we note that the dissipative stresses have at that point become comparable to the equilibrium pressure, invalidating the near-equilibrium assumption. It should be noted that this model breakdown only pertains to the purple and orange curves in Fig.~\ref{fig:orthogonal_1}.  All other curves go through their evolution without the dissipative stresses becoming too large. 

Thus we conclude that, in contrast to our investigation of the tilted case, it is possible to obtain orthogonal solutions for which the dissipative effects remain within the limits of the model throughout their evolution, and that these solutions generally remain close to their perfect fluid counterparts. On the other hand, there are solutions with large differences in the general dynamics for different $\tau$, but these are naturally accompanied by a model breakdown --- to get a large effect, the dissipative terms have to be larger than what is suitable for the near-equilibrium approach.

\section{Discussion}\label{discussion}

In this work we have rewritten the Boltzmann equation in a tetrad comoving with the equilibrium fluid. Given models for internal degrees of freedom and type of collision term, we could then 
find expressions for the thermodynamical coefficients, independent of metric. The combined Einstein-Boltzmann system was then considered. To get an as simple as possible system, but which 
contains shear viscosity and heat flow, we considered some tilted and homogeneous cosmological models with rotational symmetry. This way we obtained a self consistent system of first order ordinary differential equations, which was studied numerically for both the tilted and orthogonal cases. For orthogonal expanding models the deviations from a perfect fluid are small. However, the tilted models do not not approach an equilbrium with small tilt. Rather the tilt together with the off equilibrium quantities typically grows and the approximation of near equilibrium breaks down.

Since for realistic cosmological models the corrections to perfect fluids are most likely very small, it would be of interest to consider some denser astrophysical situations where effects of
viscosity and heat conductivity might be more prominent. In many such situations the gases are ionized, and one would have to consider more then one type of particles. The collsion term would then have to be modfied to include interactions between species.

Another approach could be to instead see the fluid as a test fluid on Schwarzschild and Kerr backgrounds, see, e.g. \cite{Kremer2} for other works along this line.

Another interesting project would be to extend the perturbative method to include terms which make the theory causal, see, e.g. \cite{Marteens,SemrenBradley}. From the point of view of the Chapman-Enskog approach to first order, it is possible to construct causal and stable theories by using the freedom of adding extra terms to $\phi_P$ that do not alter the solution up to first order \cite{Garcia-Perciante1,Garcia-Perciante2}. It would also be interesting to go to higher order in the Chapman-Enskog expansion.

\section*{Acknowlegdements}
PS and MB express their gratitude to the Centre of Mathematics at University of Minho for their kind invitation and hospitality.

\appendix

\section{Useful integrals}

On defining
\begin{equation}
I^m_n (\gamma)= \int_1^\infty\frac{(y^2-1)^m}{y^n}e^{-\gamma y} dy \ ,
\end{equation}
and assuming that $\phi(\mathcal{I}) = \mathcal{I}^\alpha$,
the integrals $A$, $B$, $C$, $D$ and $E$ can be written as
\begin{align}\nonumber
A(\gamma) 
&\equiv \frac{\gamma}{K_2(\gamma)}\int_0^\infty\frac{1}{\gamma^\star}K_2(\gamma^\star)\phi(\mathcal{I})d\mathcal{I}\\
& =\frac{\alpha!\gamma^2}{3K_2(\gamma)}\left(\frac{mc^2}{\gamma}\right)^{\alpha+1}\left[ I^{3/2}_{\alpha+1}(\gamma) + \frac{\alpha+1}{\gamma}I^{3/2}_{\alpha+2}(\gamma) \right] \label{Aint}\ ,
\end{align}
\begin{align}\nonumber
B(\gamma)&\equiv \frac{\gamma}{ K_2(\gamma)}\int_0^\infty K_3(\gamma^\star)\phi(\mathcal{I})d\mathcal{I} =\frac{\alpha!\gamma^4}{15 K_2(\gamma)}\left(\frac{mc^2}{\gamma}\right)^{\alpha+1}\times\\\
&\left[I^{5/2}_{\alpha+1}(\gamma)+\frac{3(\alpha+1)!}{\alpha!\gamma }I^{5/2}_{\alpha+2}(\gamma)+\frac{3(\alpha+2)!}{\alpha!\gamma^2 }I^{5/2}_{\alpha+3}(\gamma)+\frac{(\alpha+3)!}{\alpha!\gamma^3}I^{5/2}_{\alpha+4}(\gamma)\right] \label{Bint}\ ,
\end{align}
\begin{align}\nonumber
C(\gamma) 
&\equiv \frac{1}{K_2(\gamma)}\int_0^\infty \gamma^\star K_2(\gamma^\star)\phi(\mathcal{I})d\mathcal{I}=\frac{\alpha!\gamma^3}{3 K_2(\gamma)}\left(\frac{mc^2}{\gamma}\right)^{\alpha+1}\times\\
&\left[I^{3/2}_{\alpha+1}(\gamma)+\frac{3(\alpha+1)!}{\alpha! \gamma}I^{3/2}_{\alpha+2}(\gamma)+\frac{3(\alpha+2)!}{\alpha!\gamma^2}I^{3/2}_{\alpha+3}(\gamma)+\frac{(\alpha+3)!}{\alpha!\gamma^3}I^{3/2}_{\alpha+4}(\gamma)\right] \ ,
\end{align}
\begin{align}\nonumber
D(\gamma) 
&\equiv \frac{1}{K_2(\gamma)}\int_0^\infty \gamma^{\star 2} K_1(\gamma^\star)\phi(\mathcal{I})d\mathcal{I}=\frac{\alpha!\gamma^3}{K_2(\gamma)}\left(\frac{mc^2}{\gamma}\right)^{\alpha+1}\times\\
&\left[I^{1/2}_{\alpha+1}(\gamma)+\frac{3(\alpha+1)!}{\alpha! \gamma}I^{1/2}_{\alpha+2}(\gamma)+\frac{3(\alpha+2)!}{\alpha!\gamma^2}I^{1/2}_{\alpha+3}(\gamma)+\frac{(\alpha+3)!}{\alpha!\gamma^3}I^{1/2}_{\alpha+4}(\gamma)\right] \ ,
\end{align}
\begin{align}\nonumber
E(\gamma) &\equiv \frac{1}{K_2(\gamma)}\int_0^\infty \gamma^{\star 2} \mathrm{Ki}_1(\gamma^\star)\phi(\mathcal{I})d\mathcal{I}\\
&= \frac{\alpha!\gamma^2}{K_2(\gamma)}\left(\frac{mc^2}{\gamma}\right)^{\alpha+1}\left[I^{-1/2}_2(\gamma)+\frac{2(\alpha+1)!}{\alpha!\gamma}I^{-1/2}_3(\gamma)+\frac{(\alpha+2)!}{\alpha!\gamma^2}I^{-1/2}_4(\gamma)\right]  \ ,
\end{align}
where $\gamma^\star$ is given by (\ref{gammastar}) .

In terms of the $I^m_n$, the Bickley-Naylor functions, (\ref{BickleyNaylor}), are given by
\begin{equation}
Ki_n(\gamma)\equiv \int_1^\infty\frac{e^{-\gamma t}dt}{t^n\sqrt{t^2-1}}=I^{-1/2}_n(\gamma) \, ,
\end{equation}
and the modified Bessel functions of the second kind (\ref{BF}) by
\begin{equation}
K_n(\gamma) \equiv \left(\frac{\gamma}{2}\right)^n \frac{\Gamma(1/2)}{\Gamma(n+1/2)}
\int_{1}^{\infty} e^{-\gamma y}(y^2-1)^{n-1/2} \, dy= \left(\frac{\gamma}{2}\right)^n \frac{\Gamma(1/2)}{\Gamma(n+1/2)} I_0^{n-\frac{1}{2}}\, .
\end{equation} 

The $I_m^n$ are related by the recurrence relations
\begin{align}
\mathbf{R1} \qquad I_n^m(\gamma) &=-\frac{2m}{1-n}I^{m-1}_{n-2}(\gamma) + \frac{\gamma}{1-n}I^m_{n-1}(\gamma) \qquad n\neq 1 \ , \\
\mathbf{R2} \qquad I_n^m(\gamma) &= I_{n-2}^m(\gamma)-I_n^{m+1}(\gamma) \ .
\end{align}

Some useful relations for the Bickley-Naylor functions are given by \cite{GargantiniPomentale}
\begin{eqnarray}
nKi_{n+1}&=&(n-1)Ki_{n-1}+\gamma(Ki_{n-2}-Ki_s)\, , \\
\frac{d Ki_n}{d\gamma}&=&-Ki_{n-1} \, .
\end{eqnarray}
The following relations between the first $K_n$ and $Ki_n$ also hold
\begin{equation}
K_0=Ki_0 \quad \hbox{and} \quad K_1=Ki_1+\frac{1}{\gamma}Ki_2 \, .
\end{equation}

For completeness we also add recurrence relations for the modified Bessel functions of second kind, cf, e.g., \cite{Arfken}
\begin{eqnarray}
K_{n+1}&=&K_{n-1}+\frac{2n}{\gamma}K_n \, , \\
2\frac{d K_n}{d\gamma}&=&-K_{n-1}-K_{n+1}
\end{eqnarray}

\section{Thermodynamical coefficients}\label{AppendixCoefficient}

The energy $\bar\mu$ and the coefficients of viscosity and heat conductivity, $\bar\zeta$, $\bar\eta$ and $\bar\kappa$ are in the geometrical units given by
\begin{eqnarray}\nonumber
\bar\mu&=&n\bar{T}\left[\frac{B}{A}-1\right]\\\nonumber
\bar\zeta&=&\gamma\bar{p}\bar\tau\left[-\frac{1}{\gamma}\left(\frac{5BA-B^2+\gamma C A}{5BA-B^2+\gamma CA - A^2}\right) + \frac{B}{3\gamma A} - \frac{C-D+E}{9A}\right]=\\\nonumber
&=&-\bar{p}\bar\tau n\left(5\frac{B}{A}-\frac{B^2}{A^2}+\gamma\frac{C}{A}\right){\Large{/}}\frac{\partial\bar\mu}{\partial\bar{T}}+\frac{5}{3}\bar\eta=
-\bar{p}\bar\tau-\bar{p}\bar\tau n/\frac{\partial \bar\mu}{\partial\bar{T}}+\frac{5}{3}\bar
\eta\\\nonumber
\bar\eta&=&\frac{\gamma\bar{p}\bar\tau}{15}\left[\frac{3B}{\gamma A} - \frac{C-D+E}{A}\right]\\
\bar\kappa &=& \frac{\bar\tau \bar p}{3\bar m}\gamma^3\frac{B}{A}\left(\frac{\Gamma B}{\gamma A^2} - \frac{3}{\gamma^2}\right)\, 
\end{eqnarray}
where the integrals $A$, $B$, $C$, $D$, $E$ and $\Gamma$ are given by (\ref{intA}), (\ref{intB}), (\ref{eqC}), (\ref{intCDE}) and (\ref{Gamma}).

We now determine the derivatives $\frac{\partial\bar\mu}{\partial\bar{T}}$, $\frac{\partial\bar\zeta}{\partial\bar{T}}$ and $\frac{\partial\bar\eta}{\partial\bar{T}}$.
First we note that the dimensionless quantity $\gamma=\frac{mc^2}{kT}=\frac{\bar{m}}{\bar{T}}$. 
The derivatives of $A$, $B$, $C$, $D$ and $E$ are then given by
\begin{eqnarray}\nonumber
\frac{\partial A}{\partial\bar T}&=&\frac{B}{\bar T}-\left(\gamma\frac{K_1}{K_2}+4\right)\frac{A}{\bar T}\, , \quad
\frac{\partial B}{\partial\bar T}=\frac{\gamma}{\bar T}\left(C-\frac{K_1}{K_2}B\right)\, , \\\nonumber
\frac{\partial C}{\partial\bar T}&=&\frac{D}{\bar T}-\left(\gamma\frac{K_1}{K_2}+1\right)\frac{C}{\bar T}\, ,\\
\frac{\partial D}{\partial\bar T}&=&\frac{F}{\bar T}-\left(\gamma\frac{K_1}{K_2}+3\right)\frac{D}{\bar T}\, , \quad
\frac{\partial E}{\partial\bar T}=\frac{F}{\bar T}-\left(\gamma\frac{K_1}{K_2}+4\right)\frac{E}{\bar T}\, ,
\end{eqnarray}
where
\begin{equation}
F(\gamma)\equiv \frac{1}{K_2}\int_0^\infty\gamma^{\star 3}K_0(\gamma^\star)\Phi(\mathcal{I})d\mathcal{I} \, ,
\end{equation}
giving
\begin{equation}
\frac{\partial\bar\mu}{\partial\bar{T}}=
n\left[\frac{5B}{A}-\frac{B^2}{A^2}+\gamma\frac{C}{A}-1\right]\equiv n \bar{c}_v \, ,
\end{equation}
\begin{equation}
\frac{\partial \bar\eta}{\partial\bar T}=\frac{ n \bar\tau}{15}\frac{B}{A}\left[15-\gamma\left(\frac{D}{A}-\frac{C}{A}-\frac{E}{A}+\frac{3 B}{\gamma A}\right)\right]
\end{equation}
and
\begin{eqnarray}\nonumber
\frac{\partial \bar\zeta}{\partial\bar T}&=&\frac{n \bar\tau}{\bar{c}_v^2}
\left[1+6\gamma\frac{C}{A}+\gamma\frac{D}{A}+\frac{B}{A}\left(15-12\frac{B}{A}+2\frac{B^2}{A^2}-3\gamma\frac{C}{A}\right)\right]\\
&&-n\bar\tau+\frac{5}{3}\frac{\partial \bar\eta}{\partial\bar T}
\, ,
\end{eqnarray}
for the derivatives $\frac{\partial\bar\mu}{\partial\bar{T}}$, $\frac{\partial\bar\eta}{\partial\bar{T}}$ and $\frac{\partial\bar\zeta}{\partial\bar{T}}$ respectively.

\vskip8mm
\subsection{Monoatomic gas}
First we consider gases whose constituent particles has no internal degrees of freedom, like, e.g., monoatomic gases,
for which the integrals $A$, $B$, $C$, $D$ and $E$ simplify to (\ref{ABCDEmono}).
Hence the energy density and pressure are
\begin{equation}
\bar\mu =n \bar T\left[\gamma\frac{K_3}{K_2}-1\right]\quad \hbox{and} \quad \bar p=n\bar T
\end{equation}
and the corresponding thermodynamic coeffients, $\bar\eta$, $\bar\zeta$ and $\bar\kappa$ are
\begin{equation}
\bar\eta = \frac{n\bar T\bar\tau\gamma}{15}\left(3\frac{K_3}{K_2}-\gamma+\gamma^2\frac{K_1}{K_2}-\gamma^2\frac{Ki_1}{K_2}\right)\, ,
\end{equation}
\begin{equation}
\bar\zeta = -n\bar T\bar\tau\frac{5\gamma\frac{K_3}{K_2}-\gamma^2\frac{K_3^2}{K_2^2}+\gamma^2}{5\gamma\frac{K_3}{K_2}-\gamma^2\frac{K_3^2}{K_2^2}+\gamma^2-1}+\frac{5}{3}\bar\eta\, ,
\end{equation}
\begin{equation}
\bar\kappa = \frac{n\bar T\bar\tau}{3\bar m}\gamma^4\frac{K_3}{K_2}\left(\frac{K_3}{K_2}\left(\frac{1}{\gamma}-\frac{K_1}{K_2}+\frac{Ki_1}{K_2}\right)-\frac{3}{\gamma^2}\right)\, .
\end{equation}
Their derivatives with respect to $\bar T$ are
\begin{eqnarray}\nonumber
\frac{\partial\bar\mu}{\partial\bar T}&=&n\left(3+\gamma^2-3\gamma\frac{K_1}{K_2}-\gamma^2\frac{K_1^2}{K_2^2}\right)\\
\frac{\partial\bar\eta}{\partial\bar T}&=&-\frac{\gamma^2 n\bar\tau}{15}\left(\frac{K_1}{K_2}+\frac{4}{\gamma}\right)\left[\left(3+\gamma^2\right)\left(\frac{K_1}{K_2}-\frac{1}{\gamma}\right)
-\gamma^2\frac{Ki_1}{K_2}\right]\\\nonumber
\frac{\partial\bar\zeta}{\partial\bar T}&=&-\frac{n\bar\tau\gamma^2}{\left(\gamma x-1\right)^2}\left(\frac{\partial x}{\partial\gamma}+x^2\right)+\frac{5}{3}\frac{\partial\bar\eta}{\partial\bar{T}}\, ,
\end{eqnarray}
where
\begin{displaymath}
x\equiv -3\frac{K_1}{K_2}-\gamma\frac{K_1^2}{K_2^2}+\frac{4}{\gamma}+\gamma
\end{displaymath}
and
\begin{displaymath}
\frac{\partial x}{\partial\gamma}=3-10\frac{K_1^2}{K_2^2}-\frac{9}{\gamma}\frac{K_1}{K_2}+2\gamma\frac{K_1}{K_2}\left(1-\frac{K_1^2}{K_2^2}\right)-\frac{4}{\gamma^2}+1\, .
\end{displaymath}

\vskip8mm
\subsection{Diatomic gas}
 Next we consider gases with five internal degrees of freedom, like, e.g., diatomic gases 
for which the two vibrational degrees of freedom are excited. The adiabatic index $\alpha$ for this case equals zero, so that
$\Phi({\cal{I}})$ is constant.  The functions $A$, $B$, $C$, $D$ and $E$ then are given by equation (\ref{intABCDE2}).

The density and pressure are given by
\begin{equation}
\bar\mu=n\bar T\left(1+\gamma\frac{K_2(\gamma)}{K_1(\gamma)}\right)\quad\hbox{and}\quad \bar p=n\bar T\, ,
\end{equation}
and the thermodynamical coefficients $\bar\zeta$, $\bar\eta$ and $\bar\kappa$ by
\begin{eqnarray}\nonumber
\bar\zeta &=&n\bar T\bar\tau\left[\frac{\left(1-\gamma^2\right)}{9}-\frac{1}{1+\frac{3\gamma K_2}{K_1}+\gamma^2\left(1-\frac{K_2^2}{K_1^2}\right)}
+\frac{\gamma K_2}{9 K_1}\left(1+\frac{\gamma^2}{3}\right)+\frac{\gamma^4}{27}\left(\frac{Ki_1}{K_1}-1\right)
\right]\\
\bar\eta&=&\frac{\gamma n \bar T \bar\tau}{15}\left[\frac{10}{\gamma}-\gamma-\frac{\gamma^3}{3}\left(1-\frac{Ki_1}{K_1}\right)+\left(1+\frac{\gamma^2}{3}\right)\frac{K_2}{K_1}\right]\\
\bar\kappa&=&\frac{\gamma n\bar{T} \bar\tau}{3\bar{m}}\left(2+\gamma\frac{K_2}{K_1}\right)\left[\left(3-\gamma\frac{K_2}{K_1}+\gamma^2\left(1-\frac{Ki_1}{K_1}\right)\right)\left(2+\gamma\frac{K_2}{K_1}\right)-3\right]\, .
\end{eqnarray}
The temperature derivatives of $\bar\mu$, $\bar\eta$ and $\bar\zeta$ are
\begin{eqnarray}\nonumber
\frac{\partial\bar\mu}{\partial\bar T}&=&n\left(1+3\gamma\frac{K_2}{K_1}+\gamma^2\left(1-\frac{K^2_2}{K^2_1}\right)\right)\\
\frac{\partial\bar\eta}{\partial\bar T}&=&\frac{n\bar\tau}{15}\left(2+\gamma\frac{K_2}{K_1}\right)\left[5+\gamma\left(1+\frac{1}{3}\gamma^2\right)\left(\gamma-\frac{K_2}{K_1}\right)-\frac{\gamma^4}{3}\frac{Ki_1}{K_1}\right]\\\nonumber
\frac{\partial\bar\zeta}{\partial\bar T}&=&\frac{5}{3}\frac{\partial \bar\eta}{\partial\bar T}+\frac{n^3 \bar\tau}{\left(\frac{\partial\bar\mu}{\partial\bar T}\right)^2}
\left[\gamma\left(3-2\gamma^2\right)\frac{K_2}{K_1}+2\gamma^2\frac{K_2^2}{K_1^2}\left(\gamma\frac{K_2}{K_1}- 3\right)-1\right]-n\bar\tau\, .
\end{eqnarray}

\section{Geometric quantities in terms of metric}\label{EinsteinWeylRicci}
For the metric (\ref{BianchiVIIImetric}), the Ricci rotation coefficients are given by
\begin{eqnarray}\nonumber
\gamma^0_{\;\;10}&=&-\frac{\dot{a}}{b}\, , \quad \gamma^0_{\;\;11}=\frac{\dot{b}}{b}\, , \quad \gamma^0_{\;\;22}=\gamma^0_{\;\;33}=\frac{\dot{g}}{g}\, , \quad \gamma^0_{\;\;23}=-\gamma^0_{\;\;32}=-\frac{a}{2g^2}\, ,\\\nonumber
\gamma^1_{\;\;22}&=&\gamma^1_{\;\;33}=\frac{a\dot{g}}{bg}\, , \quad \gamma^1_{\;\;23}=-\gamma^1_{\;\;32}=-\frac{b}{2g^2}\, , \\\label{RicciRot}
\gamma^2_{\;\;30}&=&-\frac{a}{2g^2}\, , \quad \gamma^2_{\;\;31}=\frac{b}{2g^2}+\frac{1}{b} \, .
\end{eqnarray}
The nonzero components of the Einstein tensor are  
\begin{eqnarray}\nonumber
G_{00}&=&2\frac{a^2\dot g}{b^2 g}\left(\frac{\dot b}{b}-\frac{\dot a}{a}-\frac{\dot g}{2g}-\frac{\ddot g}{\dot g}\right)+\frac{3a^2}{4g^4}-\frac{b^2}{4g^4}-\frac{1}{g^2}+\frac{\dot g^2}{g^2}+2\frac{\dot b \dot g}{bg}
\\\nonumber
G_{01}&=&2\frac{a}{b}\left(\frac{\ddot g}{g}-\frac{\dot b \dot g}{bg}\right)-\frac{ab}{2g^4}
\\\nonumber
G_{11}&=&\frac{a^2}{b^2}\frac{\dot g}{g}\left(\frac{\dot g}{g}+2\frac{\dot a}{a}\right)-\frac{a^2}{4g^4}+\frac{3b^2}{4g^4}-2\frac{\ddot g}{g}-\frac{\dot g^2}{g^2}+\frac{1}{g^2}
\\\nonumber
G_{22}&=&G_{33}=\frac{a^2}{b^2}\left(\frac{\ddot a}{a}+\frac{\ddot g}{g}-\frac{\dot b}{b}\left(\frac{\dot a}{a}+\frac{\dot g}{g}\right)+\frac{\dot a}{a}\left(\frac{\dot a}{a}+2\frac{\dot g}{g}\right)\right)\\\label{EE}
&&+\frac{1}{4g^4}\left(a^2-b^2\right)-\frac{\ddot b}{b}-\frac{\ddot g}{g}-\frac{\dot b \dot g}{bg} 
\, ,
\end{eqnarray}
and the nonzero components of the electric and magnetic parts of the Weyl tensor are
 \begin{eqnarray}\label{WeylE}\nonumber
 E_{11}&=&-2E_{22}=-2E_{33}= \frac{1}{3g^4}\left(a^2-b^2-g^2\right)+\frac{1}{3}\left(\frac{\ddot{b}}{b}-\frac{\ddot{g}}{g}+\frac{\dot{g}^2}{g^2}-\frac{\dot{g}\dot{b}}{gb}\right)+\\
&& \frac{a^2}{3b^2}\left(\frac{\ddot{g}}{g}-\frac{\dot{g}^2}{g^2}-\frac{\dot{b}\dot{g}}{bg}-\frac{\ddot{a}}{a}-\frac{\dot{a}^2}{a^2}+2\frac{\dot{a}\dot{g}}{ag}+\frac{\dot{a}\dot{b}}{ab}\right)
 \end{eqnarray}
 and
 \begin{equation}\label{WeylH}
 H_{11}=-2H_{22}=-2H_{33}=\frac{a^2}{bg^2}\left(\frac{\dot{a}}{a}-\frac{\dot{g}}{g}\right)+\frac{b}{g^2}\left(\frac{\dot{g}}{g}-\frac{\dot{b}}{b}\right)\, ,
 \end{equation}
 respectively. 

\newpage
\section{Plots for the orthogonal case}\label{secPlot}
Here we show the complete solution for all quantities in the orthogonal system as discussed in section \ref{secNumOrthogonal} in the main text.
\begin{figure}[h!]
\makebox[\textwidth][c]{\includegraphics[width=1.5\textwidth]{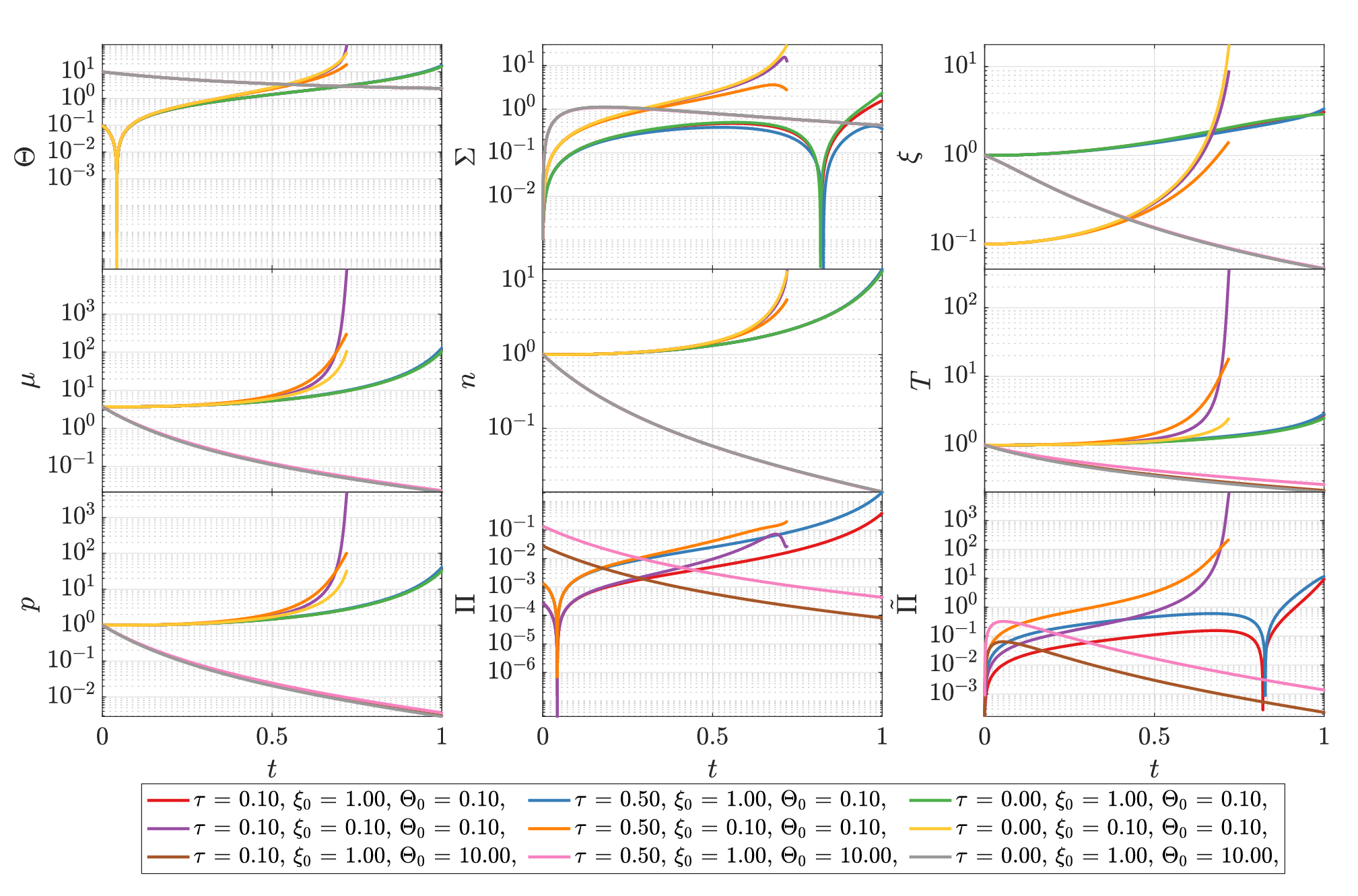}}
\caption{Solutions of the orthogonal system for different initial conditions and parameters based on (\ref{eq:TilltedICs}), but with $\mathcal{A} = \Omega = 0$.  The initial conditions and parameters that differ between the curves are indicated in the legend. Note that the quantities change sign at the cusps, but that their absolute values are given in the plots.
In particular, the diagram over $\Theta$ shows that the models of the two first lines in the legend go from an expanding to a contracting phase.}  
\label{fig:orthogonal_1_full}
\end{figure}

\end{document}